\documentclass[journal]{IEEEtran}

\usepackage{caption}
\usepackage{clrscode}
\usepackage{bbm}
\usepackage{amsfonts}
\usepackage[dvips]{graphicx}
\usepackage{times}
\usepackage{cite}
\usepackage{amsmath}
\usepackage{array}
\usepackage{amssymb}

\usepackage{stfloats}
\usepackage{slashbox}
\usepackage{graphicx}
\usepackage{subfigure}
\usepackage{footnote}
\usepackage{amsthm}
\usepackage{booktabs}
\usepackage{array}
\usepackage{algorithm}
\usepackage{algorithmic}
\usepackage{subeqnarray}
\usepackage{cases}
\usepackage{threeparttable}
\usepackage{color}
\usepackage{multirow}
\usepackage{xcolor}
\usepackage{epstopdf}
\usepackage{bm}
\usepackage{dsfont}
\usepackage{enumerate}
\usepackage{float}
\usepackage{hyperref}

\begin{document}

\title{Message-Passing Receiver Design for \\ Joint Channel Estimation and Data Decoding in Uplink Grant-Free SCMA Systems}

\author{Fan~Wei,~Wen~Chen,~\IEEEmembership{Senior~Member,~IEEE,}~Yongpeng~Wu,~\IEEEmembership{Senior~Member,~IEEE,} \\
        Jun~Ma,~Theodoros A. Tsiftsis,~\IEEEmembership{Senior~Member,~IEEE,}
\thanks{F.~Wei,~W.~Chen, and~Y.~Wu are with the Department of Electronic Engineering, Shanghai Jiao Tong University, Shanghai $200240$,China (e-mail: weifan$89$@sjtu.edu.cn; wenchen@sjtu.edu.cn; yongpeng.wu@sjtu.edu.cn).}
\thanks{J.~Ma is with School  of Electronic Engineering, Guilin University of Electronics Technology, Guangxi, China (e-mail: majun@guet.edu.cn)}
\thanks{T. A. Tsiftsis is with the School of Engineering, Nazarbayev University,
Astana $010000$, Kazakhstan (e-mail:theodoros.tsiftsis@nu.edu.kz)}
\thanks{This paper is supported by NSF China~\#$61671294$ and~\#$61671008$, National Major Project~\#$2017$ZX$03001002$-$005$, Shanghai Fundamental Key Project~\#$616$JC$1402900$, NSF Guangxi~\#$2015$GXNSFDA$139037$ and~\#$2015$GXNSFDA$139003$,and Guangxi Key Laboratory of Automatic Detecting Technology and Instruments~\#YQ$14115$.}}
\markboth{IEEE Transactions on Communications}%
{Submitted paper}

\maketitle

\begin{abstract}
The conventional grant-based network relies on the handshaking between base station and active users to achieve dynamic multi-user scheduling, which may cost large signaling overheads as well as system latency. To address those problems, the grant-free receiver design is considered in this paper based on sparse code multiple access (SCMA), one of the promising air interface technologies for $5$G wireless networks. With the presence of unknown multipath fading, the proposed receiver blindly performs joint channel estimation and data decoding without knowing the user activity in the network. Based on the framework of belief propagation (BP), we formulate a message-passing receiver for uplink SCMA that performs joint estimation iteratively. However, the direct application of BP for the multi-variable detection problem is complex. Motivated by the idea of
approximate inference, we use expectation propagation to project the intractable distributions into Gaussian families such that a linear complexity decoder is obtained. Simulation results show that the proposed receiver can detect active users in the network with a high accuracy and can achieve an improved bit-error-rate performance compared with existing methods.
\end{abstract}

\begin{IEEEkeywords}
SCMA, grant-free, user activity detection, expectation propagation, joint channel estimation and data decoding.
\end{IEEEkeywords}

\IEEEpeerreviewmaketitle

\section{Introduction}
\IEEEPARstart{W}{ith} the explosive growing demand on network capacity, throughput and connected wireless devices, the mobile broadband network is evolved into fifth generation, in which the enhanced mobile broadband (eMBB), ultra-reliable low latency communication (URLLC), and massive machine type of communication (mMTC) are three typical application scenarios. Current air interface technologies, such as orthogonal frequency-division multiple access (OFDMA) cannot fulfill the requirements in the above scenarios as orthogonal multiple access (OMA) assigns the time-frequency resources to each user exclusively. Therefore, OMA is spectrum inefficient and cannot support large throughput as well as the massive connected users in the network. In contrast, non-orthogonal multiple access (NOMA) where each resource unit shared by multiple users is supposed to be more spectrum efficient. Sparse code multiple access (SCMA)~\cite{01,02} is a code domain NOMA and is designed based on the multi-dimensional sparse signal constellation. Due to the shaping gain of multi-dimensional constellation, SCMA has a better bit-error-rate (BER) performance compared with other NOMA schemes, such as low density signature (LDS)~\cite{03}.

\subsection{Technical Literature Review}
The multi-dimensional sparse constellation forms one codebook for SCMA. In~\cite{01,02}, SCMA codebooks were heuristically designed based on the Cartesian product of quadrature amplitude modulation (QAM) symbols and followed by a unitary rotation to achieve the signal space diversity. Later, constellation rotation and interleaving was introduced for the design of multi-dimensional codebooks in order to improve the minimum Euclidean distance between SCMA codewords and combat the multipath fading~\cite{04}. To lower the peak to average power ratio (PAPR), spherical codes~\cite{05} were used to design SCMA codebooks due to the constant energy of its codes. In addition to the multi-dimensional constellation construction, factor graph matrix design was considered in~\cite{06,07} to maximize the average sum rate for uplink SCMA.

On the receiver, multiuser detector based on message-passing algorithm (MPA) was used to decode the data for each individual user. In spite of the sparsity of SCMA codewords, the decoding complexity still grows exponentially with the number of collision users in each dimension. Consequently, low complexity SCMA decoder design is of practical interests. In~\cite{02,08}, constellation with low number of projections is constructed by searching some specific unitary rotation matrices. With a reduced number of projections in each dimension, the decoding complexity for higher order SCMA codebooks is thus reduced. Apart from constellation design, the authors in~\cite{09} proposed a low complexity message-passing detector based on partial marginalization where part users are judged in advance before the message-passing iteration terminated. In~\cite{10}, Monte-Carlo Markov-Chain method was used in SCMA decoding in which the log-likelihood ratio (LLR) of each coded bit is computed based on Gibbs sampling. The Gibbs sampling has a linear complexity but also results in a slow convergence rate. Later, sphere decoder~\cite{11,12,13} was introduced in SCMA. By restricting the search scope, sphere decoder avoids exhaustive search for the possible transmitting points in each iteration. On the other hand, a low complexity decoder using expectation propagation (EP) based on the idea of approximate inference was proposed in~\cite{14}. In EP, intractable distributions are projected into some simpler families such as Gaussian distributions so that the complex computing is simplified. Moreover, low complexity hardware implementation of SCMA decoder using stochastic computing was discussed in~\cite{15}.

\subsection{Motivations}
In the above mentioned works, perfect channel state information (CSI) is assumed at receiver and the users in the network are supposed to be active simultaneously. However, this is not true in practical scenarios. Firstly, channel acquisition and tracking are always necessary and the receiver cannot obtain a perfect CSI usually. In addition, the distribution of active users in the network is always sparse in practice. In fact, according to the mobile traffic statistics~\cite{16}, the ratio of simultaneous active users in a wireless network does not exceed to $10\%$ even in the busy hours. Consequently, the base station needs to identify active users in the system before decoding their data. In LTE, the dynamic user scheduling is achieved through a request-grant procedure. However, the handshaking between base station and active users will cost the large signaling overheads as well as system latency, see~\cite{17} for the detailed discussion. In order to reduce the signaling overheads and latency, multiple access with grant-free is a promising technique in the next generation wireless networks. In~\cite{17}, the time-frequency resource referred as contention transmission unit (CTU) was defined for the uplink grant-free SCMA. Later, proof of concept (PoC) had been conducted to verify the feasibility and effectiveness of the grant-free SCMA in the user-centric no-cell (UCNC) system~\cite{18}. The active users identification and data detection for other NOMA schemes (LDS, NOMA and etc.) was studied in~\cite{19,20,21,22} based on orthogonal matching pursuit, compressive sampling matching pursuit, iterative support detection and approximate message-passing, respectively.

\subsection{Contributions}
In this paper, we focus on the design of an iterative message-passing receiver for uplink SCMA that performs joint channel estimation, data decoding, and active users detection. Iterative receivers for joint channel estimation and data decoding has been studied in~\cite{23,24,25,26,27,28,29} for MIMO-OFDM systems with the assumption that receivers have a perfect knowledge of user activity in the network. Using the framework of expectation-maximization (EM), the authors in~\cite{23,24} addressed the joint detection problem by sparse Bayesian learning (SBL) algorithm. Specifically, in joint SBL, the channel impulse responses (CIRs) are treated as hidden variables and computed in E-step while the \emph{maximum likelihood} (ML) estimation of data symbols and hyperparameters of CIRs are obtained in M-step. For multiple users or antennas system, the complexity of joint maximum in M-step grows exponentially with the number of users or antennas. Further, the decoding of data symbols in M-step belongs to hard decision which is inferior to soft detection in terms of BER performance. Factor graph (FG) and belief propagation (BP)~\cite{30} are two efficient tools to address various practical algorithms, such as forward/backward algorithm, the Viterbi algorithm, the iterative turbo decoding algorithm, Kalman filter and so on. In~\cite{25,26}, by exploring the FG structure of receivers and merging the belief propagation (BP) together with mean-field (MF) theory, the BP-MF algorithm for joint channel estimation and data detection is proposed in MIMO-OFDM system. When interferences exist, the BP-MF performs poor since it can only provide the estimate of mean values of interferences while the covariances are ignored. Moreover, accurate initialization is often required for BP-MF otherwise it will achieve a local optimal point. In~\cite{27,28,29}, Gaussian approximation in BP (BP-GA) is considered in MIMO-OFDM system via central-limit theorem and moment matching. While central-limit theorem works well in large scale MIMO, it may have a poor performance in SCMA as the collision users in one resource is limited due to the sparse structure of codewords.

In this paper, by formulating the factor graph of SCMA, we proposed a message-passing receiver based on the framework of belief propagation~\cite{30}. As the joint detection of CSI and data symbols involves a mixture of continuous and discrete variables, a concise form of BP updating rule is unavailable. Using the idea of approximate inference, we approximate the intractable distribution involved in BP with some simple families such as Gaussian distribution. However, unlike the works in~\cite{29}, the approximation does not rely on central-limit theorem. Instead, the complex distribution is projected into Gaussian families for each user individually so that the Kullback-Leibler (KL) divergence is minimized. After that, the extrinsic message for each user is updated based on expectation propagation~\cite{31}. For the active user detection, we model the CIRs for each user with the Student's t-distribution and extract the sparse signals by variational Bayesian (VB) inference. To initialize the CIRs in the algorithm, the codebook/pilot reuse is utilized so that a reliable estimation can be obtained. On the other hand, the codebook reuse can support multiple data streams for each user and therefore increase the system throughput.

In the remainder of this paper, we introduce the system model in Section II, where the factor graph is constructed for iterative message-passing receiver. In Section III-A, we consider the ML estimate of CIRs based on pilot signals only while in Section III-B, the joint channel estimation and data decoding for grant-free SCMA is discussed. The performance of our proposed receiver is evaluated in Section IV and the final conclusion is given in Section V.

\textit{Notations:} Lowercase letters $x$, bold lowercase letters $\mathbf{x}$ and bold uppercase letters $\mathbf{X}$ denote scalars, column vectors and matrices, respectively. We use $(\cdot)^{*}$ and $(\cdot)^{T}$ to denote complex conjugate, matrix transpose. $\mathcal{CN}(x;\tau,\upsilon)$ denotes the complex Gaussian distribution with mean $\tau$ and variance $\upsilon$ and $\delta(\cdot)$ denotes the Dirac delta function. $\mbox{diag}(\cdot)$ and $\mbox{Bdiag}(\cdot)$ are used to denote the diagonal matrix and block diagonal matrices respectively. $\xi \setminus k$ means the set $\xi$ with element $k$ being excluded and $\langle f(x)\rangle_{g(x)}=E_{g(x)}\{f(x)\}$.

\section{System Model}
\subsection{SCMA UL Grant-Free Transmission}
\begin{figure}
  \centering
  \includegraphics[width=3.5in]{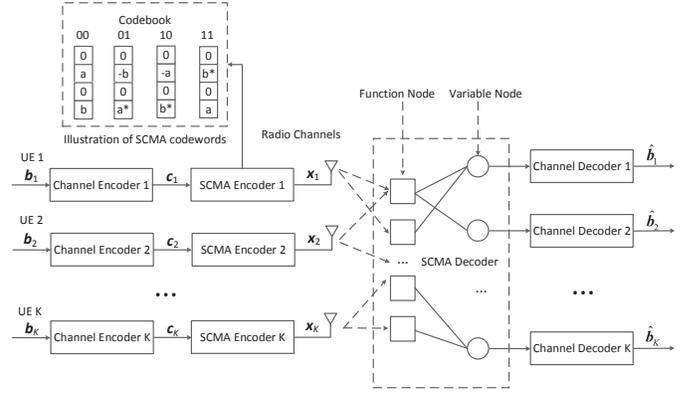}\\
  \caption{Block diagram for uplink SCMA systems}\label{Fig.1}
\end{figure}

We consider an uplink grant-free SCMA system with $K$ potential users in the network. The users can be active or inactive depending on the service requirements. In Fig.~\ref{Fig.1}, the block diagram for the uplink grant-free SCMA system is illustrated. For the active user $k$, information streams $\mathbf{b}_{k}$ are firstly sent to the channel encoder which outputs the coded bits $\mathbf{c}_{k}$. The SCMA encoder directly maps the coded bits into the multi-dimensional SCMA codeword $\mathbf{x}_{k}$ where $\mathbf{x}_{k}=(x_{1k},x_{2k},...,x_{Nk})^{T}$ is an $N$-dimensional signal constellation point. The codewords are designed to be sparse such that only $d_{v}$ of $N$ dimensions are used to transmit non-zero symbols while the remaining ones are set to be zeros. Each non-zero symbol is modulated into one OFDMA subcarrier shared by other users in the systems. Due to the sparsity of SCMA codewords, the number of overlapped users in each subcarrier $d_{c}$ is far less than $K$ so that the multiuser interferences are mitigated. At the receiver, the multiuser detector decodes the data for each active user based on received signals.

\begin{figure}
  \centering
  \includegraphics[width=3.5in]{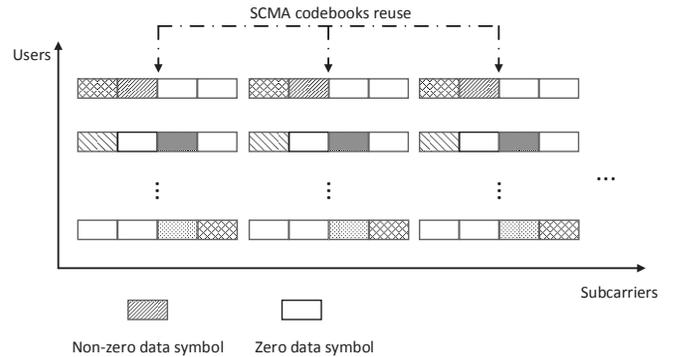}\\
  \caption{Illustration of SCMA codebook reuse}\label{Fig.2}
\end{figure}

In~\cite{17}, SCMA codebook reuse is introduced in order to support more data streams with limited resources. Based on one identical codebook, the SCMA receiver is able to detect multiple data streams as long as the codebook goes through different wireless channels. More specifically, each codebook occupies $N$ subcarriers which we call one \emph{subcarrier block} and the codebook can be reused in different \emph{subcarrier blocks}. In Fig.~\ref{Fig.2}, we illustrate the idea of codebook reuse where the X-axis denotes OFDMA subcarriers while Y-axis denotes the potential users in the system. Within this example, one subcarrier block contains $4$ OFDMA subcarriers and is reused in more than $3$ different subcarrier blocks. We assume that, in the subsequent of this paper, the SCMA codebook is reused in $B$ subcarrier blocks for each user and therefore one can transmit $B$ data streams simultaneously.

At the base station, the received signal at time slot $t$ in the $n$th subcarrier can be written as,
\begin{equation}\label{01}
  y_{tn} = \sum_{k=1}^{K} \alpha_{nk} x_{tnk} + z_{n},
\end{equation}
where $x_{tnk}$ are the transmitted signals from user $k$ which comprise both known pilot symbols and unknown SCMA codewords. In the time domain, we assume that the unknown channel coefficient $\alpha_{k}(\tau)$ can be represented by a tapped-delay model of length $L$,
\begin{equation}\label{02}
  \alpha_{k}(\tau) = \sum_{l=1}^{L} h_{kl} \delta(\tau-\tau_{l}),
\end{equation}
where $\delta(\cdot)$ is the Dirac delta function and $h_{kl}$ denotes the $l$th channel tap for user $k$. By discrete Fourier transformation (DFT), the frequency-selective block-fading channel coefficient for user $k$ in the $n$th subcarrier can be written as,
\begin{equation}\label{03}
  \alpha_{nk} = \sum_{l=1}^{L} h_{kl} \exp[-j(2\pi nl)/(BN)].
\end{equation}
Finally, $z_{n}$ is the additive complex Gaussian white noise with distribution $\mathcal{CN}(0, \sigma^{2})$. Due to the sparse structure of the SCMA codewords, the symbols $x_{tnk}$ from some users may be zero. Therefore, only $d_{c}$ instead of the whole $K$ users are collided with each other in one OFDMA subcarrier.

Writing the signals from $N$ subcarriers in a matrix form, the received signal in one subcarrier block can be written as,
\begin{equation}\label{04}
  \mathbf{y}_{t} = \sum_{k=1}^{K} \mbox{diag}(\mathbf{x}_{tk}) \mathbf{F} \mathbf{h}_{k}  + \mathbf{z},
\end{equation}
where $\mathbf{y}_{t}=(y_{t1},y_{t2},...,y_{tN})^{T}$, $\mathbf{z}=(z_{1},z_{2},...,z_{N})^{T}$ and $\mathbf{h}_{k} = (h_{k1},h_{k2},...,h_{kL})^{T}$, $\mathbf{F}$ is the $N\times L$ DFT matrix with the $(n,l)_{th}$ entry $\mathrm{F}_{nl} = \exp[-j(2\pi nl)/(BN)]$. The transmitted SCMA codewords $\mathbf{x}_{tk}=(x_{t1k},x_{t2k},...,x_{tNk})^{T}$ in each time slot are chosen randomly from a predefined alphabet set $\mathcal{X}$ with size $M$. As the user activity detection is also an interest in this paper, we introduce an augmented alphabet set $\mathcal{X}^{+} = \mathcal{X}\cup \{0\}$ with size $|\mathcal{X}^{+}|=M+1$ since an inactive user is equivalent to be transmitting zero symbols all the time.

\subsection{Factor Graph Representation}
From~\eqref{04}, the joint probability density function (PDF) of the variables is given by $p(\mathbf{C}, \mathbf{X}, \mathbf{H}, \mathbf{y})$ (time index $t$ is dropped here for notation simplification), where $\mathbf{C} = [\mathbf{c}_{1}, \mathbf{c}_{2},...,\mathbf{c}_{K}]$, $\mathbf{X} = [\mathbf{x}_{1}, \mathbf{x}_{2},...,\mathbf{x}_{K}]$ and $\mathbf{H} = [\mathbf{h}_{1}, \mathbf{h}_{2},...,\mathbf{h}_{K}]$ are the collections of coded bits, SCMA codewords and channel impulse responses (CIRs) from all users, respectively.

Based on the observation signal $\mathbf{y}$ as well as the pilot symbols, the SCMA decoder tries to find the \textit{maximum a posteriori} (MAP) estimation for each coded bit $c_{kl}$,
\begin{equation}\label{05}
  \hat{c}_{kl} = \arg \max p(c_{kl}|\mathbf{y}),
\end{equation}
where $c_{kl}$ is the $l$th coded bits for user $k$ and $p(c_{kl}|\mathbf{y})$ is given by,
\begin{equation}\label{06}
  p(c_{kl}|\mathbf{y}) \propto \sum_{\mathbf{C}\setminus c_{kl}, \mathbf{X}}\int p(\mathbf{C}, \mathbf{X}, \mathbf{H}, \mathbf{y})d\mathbf{H}.
\end{equation}
A direct computation of~\eqref{06} involves the multiple integration of continuous variables $\mathbf{H}$ and marginalization of discrete variables $\mathbf{C}$ and $\mathbf{X}$ which is prohibitively complex when the number of potential users $K$ is large. In what follows, we formulate the factor graph representation of SCMA and calculate the probability of each variable in an iterative way.

Based on the observation that $\mathbf{C}\rightarrow \mathbf{X} \rightarrow \mathbf{y}$ forms a Markov chain and CIRs $\mathbf{H}$ are independent of $\mathbf{C}$ and $\mathbf{X}$, the joint PDF can be factorized as follows,
\begin{equation}\label{07}
 p(\mathbf{C}, \mathbf{X}, \mathbf{H}, \mathbf{y}) = p(\mathbf{C})p(\mathbf{X}|\mathbf{C})p(\mathbf{y}|\mathbf{X},\mathbf{H})p(\mathbf{H}).
\end{equation}
In~\eqref{07}, $p(\mathbf{C})$ denotes the \textit{a priori} distribution of users' coded bits and $p(\mathbf{X}|\mathbf{C})$ is given by,
\begin{equation}\label{08}
 p(\mathbf{X}|\mathbf{C}) = \prod_{k}p(\mathbf{x}_{k}|\mathbf{c}_{k}),
\end{equation}
where each term $p(\mathbf{x}_{k}|\mathbf{c}_{k})$ represents the mapping function of SCMA encoder for user $k$.
Based on~\eqref{04}, the likelihood function $p(\mathbf{y}|\mathbf{X},\mathbf{H})$ can be written as,
\begin{equation}\label{09}
 p(\mathbf{y}|\mathbf{X},\mathbf{H}) \propto \exp\left\{-\frac{1}{\sigma^{2}}|\mathbf{y}
 -\sum_{k=1}^{K} \mbox{diag}(\mathbf{x}_{tk}) \mathbf{F} \mathbf{h}_{k}|^{2}\right\}.
\end{equation}

For channel $\mathbf{H}$, a non-stationary zero mean complex Gaussian \textit{priori} distribution is assumed in this paper. Due to the independence of CIRs from different users and different taps, we have,
\begin{align}\label{10}
  \nonumber p(\mathbf{H}, \pmb{\Gamma}) & = \prod_{k}q(\mathbf{h}_{k}|\pmb{\lambda}_{k}) \\
                          & = \prod_{k}\prod_{l}\mathcal{CN}(h_{kl};0,\lambda_{kl}^{-1}),
\end{align}
where $\pmb{\Gamma} = (\pmb{\lambda}_{1},\pmb{\lambda}_{2},...,\pmb{\lambda}_{K})$ and $\lambda_{kl}$ is the precision parameter and is modeled by Gamma distribution given by,
\begin{equation}\label{11}
 p(\lambda_{kl};a,b) = \mathrm{Gama}(\lambda_{kl}|a,b).
\end{equation}
After integrating the variable $\lambda_{kl}$, $p(h_{kl};a,b)$ can be written as,
\begin{align}\label{12}
  \nonumber p(h_{kl};a,b) & = \int q(h_{kl}|\lambda_{kl})p(\lambda_{kl};a,b)d\lambda_{kl} \\
                          & = \mathrm{St}(h_{kl};\mu,\nu,\gamma),
\end{align}
where the Student's t-distribution $\mathrm{St}(h_{kl};\mu,\nu,\gamma)$ is given by,
\begin{multline}\label{13}
  \mathrm{St}(h_{kl};\mu,\nu,\gamma) = \frac{\Gamma(\frac{\upsilon}{2}+1)}{\Gamma(\frac{\upsilon}{2})}\frac{2\gamma}{\pi\upsilon} \\
  \times \left[1+\frac{2\gamma}{\upsilon}|h_{kl}-\mu|^{2}\right]^{-(\frac{\upsilon}{2}+1)},
\end{multline}
with $\mu=0$, $\gamma=\frac{a}{b}$ and $\upsilon=2a$. In practice, a non-informative \textit{priori} for $a$ and $b$ is assumed and we choose $a=10^{-7}$ and $b=10^{-7}$ in this paper. The Student's t-distribution exhibits heavy tails and this property makes $h_{kl}$ favour sparse solution such that for the inactive user, most of $h_{kl}$s in $\mathbf{H}$ are near zero values. In practice, the distribution of the active users in a network is sparse and an inactive user is equivalent to have zero CIRs.

\begin{figure}
  \centering
  \includegraphics[width=3.5in,height=2.5in]{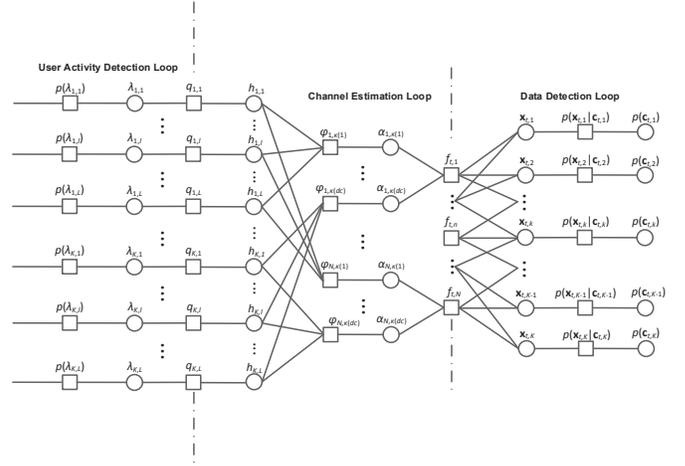}\\
  \caption{Factor graph representation of the SCMA system}\label{Fig.3}
\end{figure}

Based on the factorization~\eqref{07}, the factor graph representation of the SCMA system is plotted in Fig.~\ref{Fig.3} where circular nodes are used to denote variables (e.g., $h_{kl}$) and square nodes are used to denote functions, e.g., the function node $f_{tn}\propto \exp \left\{-\frac{1}{\sigma^{2}}|y_{tn}-\sum_{k} \alpha_{nk}x_{tnk}|^{2}\right\}$ is used to denote the likelihood function in subcarrier $n$ while constraint $\delta(\alpha_{nk}-\sum_{l=1}^{L}F_{nl}h_{kl})$ is the DFT function node $\varphi_{nk}$ for channel coefficient $\alpha_{nk}$. As can be observed in Fig.~\ref{Fig.3}, due to the sparse structure of SCMA codewords, each user choose only part of subcarriers to transmit data and only part of users are collided in each subcarrier. We define $\mathrm{F}_{n}$ to be the set of collision users in subcarrier $n$ and $\mathrm{V}_{k}$ to be the set of subcarriers for user $k$ to transmit data. In what follows, we divide the factor graph into three loops for data detection, channel estimation, and active users detection, respectively.

\section{Blind Detection for Grant-Free SCMA}
Based on the factor graph formulation in the previous section, we develop an iterative message-passing receiver for the uplink grant-free SCMA in this section. The algorithm is initialized with a pilot based channel estimation first. Thereafter, in the light of approximate inference, a low complexity iterative message-passing receiver that performs joint channel estimation, data decoding, and active users detection is proposed.
\subsection{Pilot Signals Based Channel Estimation}
In this subsection, we consider the channel estimation problem for uplink SCMA using the pilot signals received from $B$ subcarrier blocks. Assume that on the $b_{th}$ subcarrier block, the received signals at time slot $t$ can be written as,
\begin{equation}\label{14}
  \mathbf{y}_{b,t} = \sum_{k} \mathbf{X}^{p}_{b,tk} \mathbf{F}_{b} \mathbf{h}_{k} + \mathbf{n}_{b},
\end{equation}
where $\mathbf{X}^{p}_{b,tk}=\mbox{diag}(\mathbf{x}^{p}_{b,tk})$. $\mathbf{x}^{p}_{b,tk}$ is the transmitted pilots symbols that have the same sparse structure as the SCMA codewords and $\mathbf{h}_{k}$ is the CIRs for user $k$, respectively.

Let $\mathbf{y}_{t} = [\mathbf{y}_{1,t}^{T},\mathbf{y}_{2,t}^{T},...,\mathbf{y}_{B,t}^{T}]^{T}$ be the collection of received signals from $B$ subcarrier blocks, we have that,
\begin{equation}\label{15}
  \mathbf{y}_{t} = \sum_{k} \mbox{Bdiag}\{\mathbf{X}^{p}_{1,tk},\mathbf{X}^{p}_{2,tk},...,\mathbf{X}^{p}_{B,tk}\}\mathbf{\check{F}}\mathbf{h}_{k}+\mathbf{n},
\end{equation}
where $\mathbf{\check{F}}=[\mathbf{F}_{1}^{T},\mathbf{F}_{2}^{T},...,\mathbf{F}_{B}^{T}]^{T}$ is the $BN \times L$ DFT matrix. In another form, if we stack the CIRs from $K$ users into a $KL\times 1$ column vector $\mathbf{h} = [\mathbf{h}_{1}^{T},\mathbf{h}_{2}^{T},...,\mathbf{h}_{K}^{T}]^{T}$, the received signal $\mathbf{y}_{t}$ can be rewritten as,
\begin{align}\label{16}
  \nonumber \mathbf{y}_{t} & =  \begin{bmatrix}
                      \mathbf{X}_{1,t}^{p}&   &   &  \\
                      & \mathbf{X}_{2,t}^{p} &   &   \\
                      &   & \ddots &   \\
                      &   &   & \mathbf{X}_{B,t}^{p}
                     \end{bmatrix}
                     \begin{bmatrix}
                      \mbox{Bdiag}(\mathbf{F}_{1}) \\
                      \mbox{Bdiag}(\mathbf{F}_{2}) \\
                      \ldots \\
                      \mbox{Bdiag}(\mathbf{F}_{B}) \\
                     \end{bmatrix} \mathbf{h} + \mathbf{n}
  \\
  & =  \mathbf{X}_{t}\mathbf{F} \mathbf{h} + \mathbf{n},
\end{align}
where $\mathbf{X}^{p}_{b,t} = [\mathbf{X}^{p}_{b,t1},\mathbf{X}^{p}_{b,t2},...,\mathbf{X}^{p}_{b,tK}]$ is the $N \times KN$ pilot matrix and $\mbox{Bdiag}(\mathbf{F}_{b})$ is the $KN \times KL$ block diagonal matrix with the matrix on the principal diagonal being $\mathbf{F}_{b}$ for the $b_{th}$ subcarrier block, respectively.

Based on $\mathbf{y}_{t}$, the \textit{maximum likelihood} (ML) estimation of CIRs $\mathbf{h}$ is given by,
\begin{align}\label{17}
  \nonumber \mathbf{\hat{h}}_{ML} & = \arg\max\prod_{t}p(\mathbf{y}_{t}|\mathbf{h}) \\
  \nonumber  & = \arg \min \sum_{t} \|\mathbf{y}_{t}-\mathbf{X}_{t}\mathbf{F} \mathbf{h}\|^{2} \\
    & = \mathbf{G}^{-1} \mathbf{u},
\end{align}
where $\mathbf{G}=\mathbf{F}^{H}(\sum_{t}\mathbf{X}_{t}^{H}\mathbf{X}_{t})\mathbf{F}$, $\mathbf{u}=\mathbf{F}^{H}\sum_{t}\mathbf{X}_{t}^{H}\mathbf{y}_{t}$. At a first glance, the ML estimation of $\mathbf{h}$ involves a $(K\times L)$-dimensional matrix inversion so that the computational complexity grows cubically with the number of potential users $K$. In the massive connectivity scenario in which $K$ is a large number, the direct computation for~\eqref{17} would be prohibitively complex. However, if the pilot symbols from collision users are designed to be orthogonal in each subcarrier (e.g., using the Zadoff-Chu (ZC) sequence),~\eqref{17} can be computed in a low complexity alternative way.

In~\eqref{17}, the summation of the pilot symbols can be written as a block diagonal matrix,
\begin{align}\label{18}
  \nonumber \sum_{t}\mathbf{X}_{t}^{H}\mathbf{X}_{t} & = \mbox{Bdiag}\Bigg\{\sum_{t}(\mathbf{X}_{1,t}^{p})^{H}\mathbf{X}_{1,t}^{p}, \\
  &\sum_{t}(\mathbf{X}_{2,t}^{p})^{H}\mathbf{X}_{2,t}^{p},...,\sum_{t}(\mathbf{X}_{B,t}^{p})^{H}\mathbf{X}_{B,t}^{p}\Bigg\}.
\end{align}
If orthogonal pilot symbols are used such that for each subcarrier $n$,
\begin{equation}\label{19}
  \sum_{t}(x^{p}_{b,tni})^{*}x^{p}_{b,tnj} = \left\{
     \begin{array}{ll}
       1, & \hbox{$i = j$;} \\
       0, & \hbox{$i \neq j$,}
     \end{array}
   \right.
\end{equation}
$\sum_{t}(\mathbf{X}_{b,t}^{p})^{H}\mathbf{X}_{b,t}^{p}$ is also a block diagonal matrix where the $(i,j)_{th}$ submatrix is given by,
\begin{equation}\label{20}
  \sum_{t}(\mathbf{X}^{p}_{b,ti})^{H}\mathbf{X}^{p}_{b,tj} = \left\{
     \begin{array}{ll}
       \mathbf{E}_{i}, & \hbox{$i = j$;} \\
       \mathbf{0}, & \hbox{$i \neq j$.}
     \end{array}
   \right.
\end{equation}
In~\eqref{20}, $\mathbf{E}_{i}$ is a diagonal matrix with the entries on the principle diagonal being $1$ or $0$ depending on the sparse structure of SCMA codebook $i$, i.e., set $\mathrm{V}_{i}$. As a consequence, matrix $\mathbf{G}$ can be rewritten as,
\begin{equation}\label{21}
  \mathbf{G} = \sum_{b} \mbox{Bdiag}\{\mathbf{F}_{b}^{H}\mathbf{E}_{1}\mathbf{F}_{b},\mathbf{F}_{b}^{H}\mathbf{E}_{2}\mathbf{F}_{b}, \\
     ...,\mathbf{F}_{b}^{H}\mathbf{E}_{K}\mathbf{F}_{b}\}.
\end{equation}

\textit{Proposition 1:} A necessary condition for matrix $\mathbf{G}$ to be invertible is $Bd_{v}\geq L$.

\textit{Proof:} From~\eqref{21}, matrix $\mathbf{G}$ reduced to a block diagonal matrix. To get the inverse matrix of $\mathbf{G}$, we must compute the inverse matrix of sub-block $\mathbf{F}_{b}^{'}=\sum_{b}\mathbf{F}_{b}^{H}\mathbf{E}_{k}\mathbf{F}_{b}$, i.e., the $k_{th}$ submatrix of $\mathbf{G}$. Note that the sparse matrix $\mathbf{E}_{k}$ has a rank of $d_{v}$ and $\mathbf{F}_{b}^{'}$ is an $L$-dimensional square matrix. As a result, the necessary condition for matrix $\mathbf{G}$ to be invertible is $\mathrm{rank}(\mathbf{F}_{b}^{'})=B\min\{L,d_{v}\}\geq L$. $\square$

From the Proposition $1$, one can infer that to get a convinced estimation of $\mathbf{h}$, the number of nonzero subcarriers for each user transmitting pilot symbols must exceed the number of channel taps. Further, since we only need to compute the inverse of each submatrix in~\eqref{21}, the computational complexity thus reduces to the order of $\mathcal{O}(KL^{3})$, which grows linearly with the number of potential users.

\subsection{Joint Channel Estimation and Data Decoding}
In this subsection, we discuss the joint detection methods based on transmitted pilot as well as data symbols. Recall that in Section II, we have divided the factor graph of SCMA into three loops as shown in Fig~\ref{Fig.3}. Based on the FG of SCMA, the message-passing formulas for the three loops are developed.
\subsubsection{Joint detection for data decoding}
We begin our discussion with the data detection loop first. In the previous section, we have analyzed the direct MAP estimation, which involves a mixture of continuous and discrete variables is of high complexity. Consequently, to reduce the detection complexity, we should obtain the joint estimation in a low complexity iterative way. For the systems with the FG structure, belief propagation is an iterative algorithm that deals with the multi-variables problems. Based on the BP updating rule, the message sent from function node $f_{tn}$ to variable node $\mathbf{x}_{tk}$ can be written as (we drop the subscript $b$ here for notation simplification),
\begin{multline}\label{22}
  I_{f_{tn}\rightarrow\mathbf{x}_{tk}}(x_{tnk}) = \sum_{\mathbf{x}_{i}:i\in \mathrm{F}_{n}\setminus k}
  I_{\mathbf{x}_{ti}\rightarrow f_{tn}}(x_{tni}) \\
  \cdot \int f_{tn}(\mathbf{X}_{tn}, \pmb{\alpha}_{n})\prod_{i\in \mathrm{F}_{n}}
  I_{\alpha_{ni}\rightarrow f_{tn}}(\alpha_{ni})d\pmb{\alpha_{n}},
\end{multline}
where $I_{\mathbf{x}_{ti}\rightarrow f_{tn}}(x_{tni})$ and $I_{\alpha_{ni}\rightarrow f_{tn}}(\alpha_{ni})$ are the extrinsic message passed from nodes $\mathbf{x}_{ti}$ and $\alpha_{ni}$ to $f_{tn}$, respectively. The likelihood function $f_{tn}(\mathbf{X}_{tn}, \pmb{\alpha}_{n})$ is given by,
\begin{equation}\label{23}
  f_{tn}(\mathbf{X}_{tn}, \pmb{\alpha}_{n}) \propto \exp \left\{-\frac{1}{\sigma^{2}}|y_{tn}-\sum_{k\in \mathrm{F}_{n}} \alpha_{nk}
  x_{tnk}|^{2}\right\},
\end{equation}
where $\mathbf{X}_{tn}$ and $\pmb{\alpha}_{n}$ denote the transmitted data symbols and CIRs from collision users in subcarrier $n$, respectively.

From~\eqref{22}-\eqref{23}, one can observe that the BP algorithm has reduced the computational scale for joint estimation from $K$ users to $d_{c}$ users within each iteration. However, as the function node $f_{tn}$ involves a mixture of discrete variables $x_{tnk}$ and continuous variables $\alpha_{nk}$, direct computation of~\eqref{22} is still of high complexity due to the multiple integration as well as multiple marginalization.

To reduce the computational complexity of~\eqref{22}, the belief propagation mean-field message passing is proposed based on variational Bayesian (VB) inference~\cite{25,26,32}. Specifically, with BP-MF, the message-passing for $x_{tnk}$ can be updated as,
\begin{align}\label{23a}
  \nonumber \ln I_{f_{tn}\rightarrow \mathbf{x}_{tk}} (x_{tnk}) & = \sum_{u \in \mathrm{F}_{n}\setminus k}
            b(\mathbf{x}_{tu}) \\
  \nonumber \cdot \int \ln & f_{tn}(\mathbf{X}_{n},\pmb{\alpha}_{n})\prod_{u\in \mathrm{F}_{n}}b(\alpha_{nu}) \prod_{u\in
            \mathrm{F}_{n}}d\alpha_{nu} \\
            & \propto \ln \mathcal{CN}(x_{tnk}; \tau_{f_{tn}\rightarrow \mathbf{x}_{tk}}, \upsilon_{f_{tn}\rightarrow
            \mathbf{x}_{tk}}),
\end{align}
where $b(\mathbf{x}_{tu})$ and $b(\alpha_{nu})$ is the posterior belief~\cite{30} of variable $\mathbf{x}_{tu}$ and $\alpha_{nu}$, respectively and $\tau_{f_{tn}\rightarrow \mathbf{x}_{tk}}$, $\upsilon_{f_{tn}\rightarrow\mathbf{x}_{tk}}$ are given by,
\begin{align}\label{23b}
  & \tau_{f_{tn}\rightarrow \mathbf{x}_{tk}} = \frac{\tau_{\alpha_{nk}}^{*}(y_{tn}
  -\sum_{u\in \mathbf{F}_{n}\backslash k}\tau_{\alpha_{nu}}\tau_{x_{tnu}})}{|\tau_{\alpha_{nk}}|^{2}+\upsilon_{\alpha_{nk}}}, \\
  & \upsilon_{f_{tn}\rightarrow \mathbf{x}_{tk}} = \sigma^{2}(|\tau_{\alpha_{nk}}|^{2}+\upsilon_{\alpha_{nk}})^{-1},
\end{align}
where $\tau_{\alpha_{nk}}$ and $\upsilon_{\alpha_{nk}}$ are the mean and covariance of $\alpha_{nk}$ with respect to belief $b(\alpha_{nk})$. Similarly, $\tau_{x_{tnk}}$ is the mean of $x_{tnk}$ with respect to belief $b(\mathbf{x}_{tk})$.
Compared with BP updating, MF has a simple updating rule, in particular for conjugate-exponential models. However, for multiuser system, the interference cancellation structure in~\eqref{23b} only involves the mean values of interferences while the covariances are not being considered. This makes BP-MF perform poor in the estimation of LLRs for data symbols. In an alternative way, the interferences are modeled as Gaussian distributions based on central-limit theorem~\cite{27,28,29}. While central-limit theorem is effective in the massive MIMO-OFDM system, it may result in a large performance degradation in SCMA since the number of collision users in each subcarrier is limited ($d_{c}$ compared with the total number of users $K$) due to the sparse structure of SCMA codewords.

In this paper, instead of using central-limit theorem, the distribution of each interference $\alpha_{nk}x_{tnk}$ is projected into Gaussian families separately using expectation propagation algorithm. Based on the idea of divide-and-conquer, EP can be viewed as a distributed inference where the big data is partitioned into smaller pieces and local inference is performed for each one separately. Meanwhile, expectation propagation belongs to a class of approximate inference that the intractable distributions are always approximated to some simpler distributions (e.g., Gaussian distributions) by minimizing the Kullback-Leibler (KL) divergence,
\begin{equation}\label{24}
  D_{KL}(p(x)||q(x)) = \int p(x)\log\frac{p(x)}{q(x)}dx,
\end{equation}
where $p(x)$ is the true distribution and $q(x)$ is the approximated distribution.

\textit{Proposition 2:} Let $u_{tnk} = \alpha_{nk}x_{tnk}$, given the extrinsic messages $I_{\mathbf{x}_{tk}\rightarrow f_{tn}}(x_{tnk})$ and $I_{\alpha_{nk}\rightarrow f_{tn}}(\alpha_{nk})$ follows Gaussian distribution $\mathcal{CN}(\alpha_{nk}; \tau_{\alpha_{nk}\rightarrow f_{tn}},\upsilon_{\alpha_{nk}\rightarrow f_{tn}})$, the probability density function (PDF) of variable $u_{tnk}$ can be written by~\eqref{27} shown at the top of next page.

\textit{Proof :} See Appendix.
\begin{figure*}[t]
  \begin{equation}\label{27}
   I_{u_{tnk}\rightarrow f_{tn}}(u_{tnk}) \propto
   \left\{
     \begin{array}{ll}
       \sum_{x_{tnk}}I_{\mathbf{x}_{t,k}\rightarrow f_{tnk}}(x_{tnk})|x_{tnk}|\mathcal{CN}(u_{tnk};\tau_{\alpha_{nk}\rightarrow
       f_{tnk}}x_{tnk}, \upsilon_{\alpha_{nk}\rightarrow f_{tnk}} |x_{tnk}|^{2}), & \hbox{$x_{tnk}\neq 0$;} \\
       I_{\mathbf{x}_{tk} \rightarrow f_{tn}}(x_{tnk}), & \hbox{$x_{tnk} = 0$.}
     \end{array}
   \right.
  \end{equation}
  \begin{align}\label{28}
  \nonumber b(u_{tnk}) &= I_{u_{tnk}\rightarrow f_{tn}}(u_{tnk})I_{f_{tn} \rightarrow u_{tnk}}(u_{tnk})\\
           & = \left\{
              \begin{array}{ll}
               \sum_{x_{tnk}}\beta(x_{tnk})\mathcal{CN}(u_{tnk};\tau_{u_{tnk}},\upsilon_{u_{tnk}}), &\hbox{$x_{tnk}\neq 0$;} \\
                C^{-1}I_{\mathbf{x}_{tk} \rightarrow f_{tn}}(x_{tnk})\mathcal{CN}(u_{tnk};\tau_{f_{tn}\rightarrow
                u_{tnk}},\upsilon_{f_{tn}\rightarrow u_{tnk}}), & \hbox{$x_{tnk}=0$.}
              \end{array}
             \right.
  \end{align}
  \begin{equation}\label{29}
    \beta(x_{tnk})= C^{-1}I_{\mathbf{x}_{t,k}\rightarrow f_{tnk}}(x_{tnk})|x_{tnk}|\mathcal{CN}(\tau_{f_{tn} \rightarrow
    u_{tnk}};\tau_{\alpha_{nk}\rightarrow f_{tnk}}x_{tnk},\upsilon_{f_{tn} \rightarrow u_{tnk}}+\upsilon_{\alpha_{nk}\rightarrow f_{tnk}}|x_{tnk}|^{2}),
  \end{equation}
  \begin{equation}\label{30}
  \tau_{u_{tnk}}
   = \frac{\tau_{\alpha_{nk}\rightarrow f_{tnk}}\upsilon_{f_{tn} \rightarrow u_{tnk}}x_{tnk}+\tau_{f_{tn} \rightarrow
   u_{tnk}}\upsilon_{\alpha_{nk}\rightarrow f_{tnk}}|x_{tnk}|^{2}}
   {\upsilon_{f_{tn} \rightarrow u_{tnk}}+\upsilon_{\alpha_{nk}\rightarrow f_{tnk}}|x_{tnk}|^{2}},
  \end{equation}
  \hrulefill
\end{figure*}

As can be seen from~\eqref{27}, the probability density function of $u_{tnk}$ is a mixture of Gaussian functions and is discontinuous at $u_{tnk}=0$. This makes the BP message passing involves complicated computing in each iteration. In order to reduce the computational complexity, we resort to EP method to project $I_{u_{tnk}\rightarrow f_{tn}}(u_{tnk})$ into Gaussian distribution. Note that in many cases, there is only one significant component contributes to the Gaussian mixture, thus Gaussian distribution suffices to be a good approximation for~\eqref{27}.

Assume that the output information $I_{f_{tn} \rightarrow u_{tnk}}(u_{tnk})$ from function node $f_{tn}$ follows the Gaussian distribution $\mathcal{CN}(u_{tnk};\tau_{f_{tn}\rightarrow u_{tnk}},\upsilon_{f_{tn}\rightarrow u_{tnk}}) $, the posterior belief of variable $u_{tnk}$ can be calculated in~\eqref{28} where $\beta(x_{tnk})$, $\tau_{u_{tnk}}$ and $\upsilon_{u_{tnk}}$ are computed in~\eqref{29}-\eqref{31} and $C$ is a normalization constant.
\begin{equation}\label{31}
   \upsilon_{u_{tnk}}
   =\frac{\upsilon_{f_{tn} \rightarrow u_{tnk}}\upsilon_{\alpha_{nk}\rightarrow f_{tnk}}|x_{tnk}|^{2}}
   {\upsilon_{f_{tn} \rightarrow u_{tnk}}+\upsilon_{\alpha_{nk}\rightarrow f_{tnk}}|x_{tnk}|^{2}}.
\end{equation}
Notice that in~\eqref{28}, the posterior belief of variable $u_{tnk}$ is comprising of two pieces. Based on EP, local inference is proceeded for message of $I_{u_{tnk}\rightarrow f_{tn}}(u_{tnk})$. To start with, $b(u_{tnk})$ is projected into a Gaussian distribution $\hat{b}(u_{tnk})$ so that the KL-divergence $D_{KL}[b(u_{tnk})||\hat{b}(u_{tnk})]$ is minimized. The result is reduced to moment matching such that,
\begin{equation}\label{32}
  \hat{b}(u_{tnk})= \mathcal{CN}(u_{tnk};\hat{\tau}_{u_{tnk}}, \hat{\upsilon}_{u_{tnk}}),
\end{equation}
\begin{equation}\label{33}
  \hat{\tau}_{u_{tnk}} = \sum_{x_{tnk}} \beta(x_{tnk})\tau_{u_{tnk}},
\end{equation}
\begin{equation}\label{34}
 \hat{\upsilon}_{u_{tnk}} = \sum_{x_{tnk}}
 \beta(x_{tnk})(|\tau_{u_{tnk}}|^{2}+\upsilon_{u_{tnk}})-|\hat{\tau}_{u_{tnk}}|^{2}.
\end{equation}
Assume $I_{f_{tn} \rightarrow u_{tnk}}(u_{tnk})$ follows Gaussian distribution, by EP principle, we have
\begin{align}\label{35}
  \nonumber I_{u_{tnk}\rightarrow f_{tn}}(u_{tnk}) &= \frac{\hat{b}(u_{tnk})}{I_{f_{tn} \rightarrow u_{tnk}}(u_{tnk})} \\
  & \propto \mathcal{CN}(u_{tnk};\tau_{u_{tnk}\rightarrow f_{tn}},\upsilon_{u_{tnk}\rightarrow f_{tn}}),
\end{align}
where
\begin{equation}\label{36}
  \tau_{u_{tnk}\rightarrow f_{tn}} = \upsilon_{u_{tnk}\rightarrow f_{tn}}
  \left(\frac{\hat{\tau}_{u_{tnk}}}{\hat{\upsilon}_{u_{tnk}}}
  -\frac{\tau_{f_{tn} \rightarrow u_{tnk}}}{\upsilon_{f_{tn} \rightarrow u_{tnk}}}\right),
\end{equation}
\begin{equation}\label{37}
  \upsilon_{u_{tnk}\rightarrow f_{tn}}= \left(\frac{1}{\hat{\upsilon}_{u_{tnk}}}
  -\frac{1}{\upsilon_{f_{tn} \rightarrow u_{tnk}}}\right)^{-1}.
\end{equation}
Now given the extrinsic messages of interferences $u_{tni},i \in \mathrm{F}_{n}\backslash k$ follow independent Gaussian distributions, $u_{tnk} = y_{tn} - \sum_{i\in \mathbf{F}_{n}\setminus k} u_{tni}$ is also a Gaussian variable such that,
\begin{equation}\label{38}
  I_{f_{tn} \rightarrow u_{tnk}}(u_{tnk}) \sim
  \mathcal{CN}(u_{tnk};\tau_{f_{tn}\rightarrow u_{tnk}},\upsilon_{f_{tn}\rightarrow u_{tnk}}),
\end{equation}
where
\begin{equation}\label{39}
  \tau_{f_{tn} \rightarrow u_{tnk}} = y_{tn} - \sum_{i\in \mathrm{F}_{n}\setminus k} \tau_{u_{tni}\rightarrow f_{tn}},
\end{equation}
\begin{equation}\label{40}
  \upsilon_{f_{tn} \rightarrow u_{tnk}} =\sigma^{2}+\sum_{i\in \mathrm{F}_{n}\setminus k}
  \upsilon_{u_{tni}\rightarrow f_{tn}}.
\end{equation}

Since $ I_{f_{tn} \rightarrow u_{tnk}}(u_{tnk})$ follows Gaussian distribution, the message sent from function node $f_{tn}$ to variable node $\mathbf{x}_{tk}$ can be updated as,
\begin{align}\label{41}
  \nonumber I_{f_{tn}\rightarrow \mathbf{x}_{tk}}(x_{tnk}) & = \int I_{f_{tn} \rightarrow u_{tnk}}(u_{tnk})I_{\alpha_{nk}\rightarrow f_{tn}}(\alpha_{nk}) d\alpha_{nk} \\
  & \propto \exp\{-\Delta_{f_{tn}\rightarrow x_{tnk}}(x_{tnk})\},
\end{align}
where
\begin{align}\label{42}
  \nonumber \Delta_{f_{tn}\rightarrow x_{tnk}}(x_{tnk})
  & = \frac{|\tau_{f_{tn}\rightarrow u_{tnk}}-\tau_{\alpha_{nk}\rightarrow f_{tn}}x_{tnk}|^{2}}{\upsilon_{f_{tn}\rightarrow
    u_{tnk}}+\upsilon_{\alpha_{nk}\rightarrow f_{tn}}|x_{tnk}|^{2}}\\
  & + \ln(\upsilon_{f_{tn}\rightarrow u_{tnk}}+\upsilon_{\alpha_{nk}\rightarrow f_{tn}}|x_{tnk}|^{2}),
\end{align}
and we have assumed that the message $I_{\alpha_{nk}\rightarrow f_{tn}}(\alpha_{nk})$ follows $\mathcal{CN}(\alpha_{nk};\tau_{\alpha_{nk}\rightarrow f_{tn}},\upsilon_{\alpha_{nk}\rightarrow f_{tn}})$ which will be given in~\eqref{67}. Notice that in computing~\eqref{42}, the LLRs for data symbols, not only the mean values but also the covariances of interferences are involved.

With $I_{f_{tm}\rightarrow \mathbf{x}_{tk}}(x_{tnk})$ computed in the previous step, the message $I_{\mathbf{x}_{tk} \rightarrow f_{tn}}(x_{tnk})$ can be calculated according to the BP rule,
\begin{align}\label{43}
  \nonumber I_{\mathbf{x}_{tk} \rightarrow f_{tn}}(x_{tnk}) &= p(\mathbf{x}_{tk}|\mathbf{c}_{tk})\prod_{m\neq n}I_{f_{tm}\rightarrow
            \mathbf{x}_{tk}}(x_{tmk}) \\
  & \propto \exp\Bigg\{-\sum_{m\in \mathrm{V}_{k} \backslash n}\Delta_{f_{tm}\rightarrow x_{tmk}}(x_{tmk})\Bigg\}.
\end{align}
\subsubsection{Data aided channel estimation and user activity detection}
Now we begin our discussion on the channel estimation loop in Fig.~\ref{Fig.3}. As in~\eqref{41}, with $I_{f_{tn} \rightarrow u_{tnk}}(u_{tnk})$ follows Gaussian distribution, the message passed from function node $f_{tn}$ to the variable node $\alpha_{nk}$ can be updated as,
\begin{align}\label{44}
  \nonumber I_{f_{tn} \rightarrow \alpha_{nk}}(\alpha_{nk}) & = \sum_{x_{tnk}} I_{\mathbf{x}_{tk} \rightarrow f_{tn}}(x_{tnk}) I_{f_{tn} \rightarrow u_{tnk}}(u_{tnk}) \\
  \nonumber & =\sum_{x_{tnk}}I_{\mathbf{x}_{tk} \rightarrow f_{tn}}(x_{tnk}) \\
   & \cdot \mathcal{CN}(\alpha_{nk}x_{tnk};\tau_{f_{tn}\rightarrow
      u_{tnk}},\upsilon_{f_{tn}\rightarrow u_{tnk}}),
\end{align}
which is a mixture of Gaussian distributions. To get a simpler form for $I_{f_{tn} \rightarrow \alpha_{nk}}(\alpha_{nk})$, again we project $I_{f_{tn} \rightarrow \alpha_{nk}}(\alpha_{nk})$ into Gaussian families by EP.

The posterior belief of $\alpha_{nk}$ is given by,
\begin{align}\label{45}
  \nonumber b(\alpha_{nk}) &= I_{f_{tn} \rightarrow \alpha_{nk}}(\alpha_{nk})I_{\alpha_{nk} \rightarrow f_{tn}}(\alpha_{nk})  \\
  & = \sum_{x_{tnk}}\beta(x_{tnk})\mathcal{CN}(\alpha_{nk};\tilde{\tau}_{\alpha_{nk}},\tilde{\upsilon}_{\alpha_{nk}}),
\end{align}
where we have assumed $I_{\alpha_{nk} \rightarrow f_{tn}}(\alpha_{nk})$ follows a Gaussian distribution. $\beta(x_{tnk})$, $\tilde{\tau}_{\alpha_{nk}}$ and $\tilde{\upsilon}_{\alpha_{nk}}$ are computed in~\eqref{46}-\eqref{48} and $C$ is a normalization constant.
\begin{figure*}[t]
  \begin{equation}\label{46}
    \beta(x_{tnk}) = C^{-1}I_{\mathbf{x}_{tk} \rightarrow f_{tn}}(x_{tnk})
    \mathcal{CN}(\tau_{\alpha_{nk} \rightarrow f_{tn}}x_{tnk};\tau_{f_{tn}\rightarrow u_{tnk}},\upsilon_{f_{tn}\rightarrow u_{tnk}}
    +\upsilon_{\alpha_{nk} \rightarrow f_{tn}}|x_{tnk}|^2),
  \end{equation}
  \begin{equation}\label{47}
    \tilde{\tau}_{\alpha_{nk}} = \frac{\tau_{\alpha_{nk} \rightarrow f_{tn}}\upsilon_{f_{tn}\rightarrow u_{tnk}}+\tau_{f_{tn}\rightarrow u_{tnk}} \upsilon_{\alpha_{nk} \rightarrow f_{tn}}x_{tnk}^{*}}
    {\upsilon_{f_{tn}\rightarrow u_{tnk}}+\upsilon_{\alpha_{nk} \rightarrow f_{tn}}|x_{tnk}|^2},
  \end{equation}
  \hrulefill
\end{figure*}
\begin{equation}\label{48}
  \tilde{\upsilon}_{\alpha_{nk}} = \frac{\upsilon_{f_{tn}\rightarrow u_{tnk}}\upsilon_{\alpha_{nk} \rightarrow f_{tn}}}
  {\upsilon_{f_{tn}\rightarrow u_{tnk}}+\upsilon_{\alpha_{nk} \rightarrow f_{tn}}|x_{tnk}|^2}.
\end{equation}
We project $b(\alpha_{nk})$ into Gaussian distribution so that the KL-divergence $D_{KL}[b(\alpha_{nk})||\hat{b}(\alpha_{nk})]$ is minimized,
\begin{equation}\label{49}
  \hat{b}(\alpha_{nk}) = \mathcal{CN}(\alpha_{nk};\hat{\tau}_{\alpha_{nk}},\hat{\upsilon}_{\alpha_{nk}}),
\end{equation}
where by moment matching,
\begin{equation}\label{50}
  \hat{\tau}_{\alpha_{nk}} = \sum_{x_{tnk}}\beta(x_{tnk})\tilde{\tau}_{\alpha_{nk}},
\end{equation}
\begin{equation}\label{51}
  \hat{\upsilon}_{\alpha_{nk}} = \sum_{x_{tnk}}\beta(x_{tnk})(|\tilde{\tau}_{\alpha_{nk}}|^{2}+\tilde{\upsilon}_{\alpha_{nk}})-
  |\hat{\tau}_{\alpha_{nk}}|^{2}.
\end{equation}
Since the extrinsic message $I_{\alpha_{nk} \rightarrow f_{tn}}(\alpha_{nk})$ follows a Gaussian distribution $\mathcal{CN}(\alpha_{nk};\tau_{\alpha_{nk} \rightarrow f_{tn}},\upsilon_{\alpha_{nk} \rightarrow f_{tn}})$, by EP principle, we have,
\begin{align}\label{52}
  \nonumber I_{f_{tn} \rightarrow \alpha_{nk}}(\alpha_{nk})&=\frac{\hat{b}(\alpha_{nk})}{I_{\alpha_{nk} \rightarrow f_{tn}}(\alpha_{nk})} \\
  & \propto \mathcal{CN}(\alpha_{nk};\tau_{f_{tn} \rightarrow \alpha_{nk}},\upsilon_{f_{tn} \rightarrow \alpha_{nk}}),
\end{align}
where
\begin{equation}\label{53}
  \tau_{f_{tn} \rightarrow \alpha_{nk}} = \upsilon_{f_{tn} \rightarrow \alpha_{nk}}
  \left(\frac{\hat{\tau}_{\alpha_{nk}}}{\hat{\upsilon}_{\alpha_{nk}}}-\frac{\tau_{\alpha_{nk} \rightarrow f_{tn}}}{\upsilon_{\alpha_{nk} \rightarrow f_{tn}}}\right),
\end{equation}
\begin{equation}\label{54}
  \upsilon_{f_{tn} \rightarrow \alpha_{nk}} = \left(\frac{1}{\hat{\upsilon}_{\alpha_{nk}}}-\frac{1}{\upsilon_{\alpha_{nk} \rightarrow f_{tn}}}\right)^{-1}.
\end{equation}
For pilot signals,~\eqref{52} is reduced to,
\begin{align}\label{55}
  \nonumber I_{f_{tn} \rightarrow \alpha_{nk}}(\alpha_{nk}) & = \mathcal{CN}(\alpha_{nk}x_{tnk}^{p};\tau_{f_{tn} \rightarrow u_{tnk}},\upsilon_{f_{tn} \rightarrow u_{tnk}}), \\
    & \propto \mathcal{CN}(\alpha_{nk};\tau_{f_{tn} \rightarrow \alpha_{nk}},\upsilon_{f_{tn} \rightarrow \alpha_{nk}}),
\end{align}
where
\begin{equation}\label{56}
  \tau_{f_{tn} \rightarrow \alpha_{nk}} = \frac{\tau_{f_{tn} \rightarrow u_{tnk}}}{x_{tnk}^{p}},
\end{equation}
\begin{equation}\label{57}
  \upsilon_{f_{tn} \rightarrow \alpha_{nk}} = \frac{\upsilon_{f_{tn} \rightarrow u_{tnk}}}{|x_{tnk}^{p}|^{2}}.
\end{equation}

Now for both pilot and data signals, $I_{f_{tn} \rightarrow \alpha_{nk}}(\alpha_{nk})$ follows the Gaussian distributions. According to the BP updating rule, we have,
\begin{align}\label{58}
  \nonumber I_{\alpha_{nk} \rightarrow \varphi_{nk} }(\alpha_{nk}) & = \prod_{t} I_{f_{tn}\rightarrow \alpha_{nk}}(\alpha_{nk})  \\
  & \propto \mathcal{CN}(\alpha_{nk}; \tau_{\alpha_{nk} \rightarrow \varphi_{nk}}, \upsilon_{\alpha_{nk} \rightarrow \varphi_{nk}}),
\end{align}
where
\begin{equation}\label{59}
  \tau_{\alpha_{nk} \rightarrow \varphi_{nk}} = \upsilon_{\alpha_{nk} \rightarrow \varphi_{nk}}
  \sum_{t} \frac{\tau_{f_{tn}\rightarrow \alpha_{nk}}}{\upsilon_{f_{tn}\rightarrow \alpha_{nk}}},
\end{equation}
\begin{equation}\label{60}
  \upsilon_{\alpha_{nk} \rightarrow \varphi_{nk}} = \left(\sum_{t} \frac{1}{\upsilon_{f_{tn}\rightarrow \alpha_{nk}}}\right)^{-1},
\end{equation}
and the production in~\eqref{58} is through all time slots that the block fading coefficient $\alpha_{nk}$ maintains unchanged.

As for the message updating in the reverse direction, $I_{\varphi_{nk} \rightarrow \alpha_{nk}}(\alpha_{nk})$ can be calculated with the assumption that $I_{h_{kl} \rightarrow \varphi_{nk}}(h_{kl})$ follows Gaussian distribution shown in~\eqref{79},
\begin{align}\label{61}
 \nonumber I_{\varphi_{nk} \rightarrow \alpha_{nk}}(\alpha_{nk}) & = \int \varphi_{nk}(\alpha_{nk},\mathbf{h}_{k}) \prod_{l}
  I_{h_{kl} \rightarrow \varphi_{nk}}(h_{kl})d\mathbf{h}_{k} \\
  & \propto \mathcal{CN}(\alpha_{nk};\tau_{\varphi_{nk} \rightarrow \alpha_{nk}},\upsilon_{\varphi_{nk} \rightarrow \alpha_{nk}}),
\end{align}
where $\varphi_{nk}(\alpha_{nk},\mathbf{h}_{k}) = \delta(\alpha_{nk}-\sum_{l}\mathrm{F}_{nl}h_{kl})$ and
\begin{equation}\label{62}
  \tau_{\varphi_{nk} \rightarrow \alpha_{nk}} = \sum_{l} \mathrm{F}_{nl} \tau_{h_{kl} \rightarrow \varphi_{nk}},
\end{equation}
\begin{equation}\label{63}
  \upsilon_{\varphi_{nk} \rightarrow \alpha_{nk}} = \sum_{l} \mathrm{F}_{nl}^{2} \upsilon_{h_{kl} \rightarrow \varphi_{nk}}.
\end{equation}
With~\eqref{58} and~\eqref{61}, the posterior belief of $\alpha_{nk}$ can be calculated as,
\begin{align}\label{64}
 \nonumber b(\alpha_{nk}) & = I_{\alpha_{nk}\rightarrow \varphi_{nk}}(\alpha_{nk}) I_{\varphi_{nk} \rightarrow \alpha_{nk}}(\alpha_{nk}) \\
  & \propto \mathcal{CN}(\alpha_{nk}; \tau_{\alpha_{nk}}, \upsilon_{\alpha_{nk}}),
\end{align}
where
\begin{equation}\label{65}
  \tau_{\alpha_{nk}} = \frac{\tau_{\alpha_{nk}\rightarrow \varphi_{nk}}\upsilon_{\varphi_{nk} \rightarrow \alpha_{nk}}
  +\tau_{\varphi_{nk} \rightarrow \alpha_{nk}}\upsilon_{\alpha_{nk}\rightarrow \varphi_{nk}}}
  {\upsilon_{\varphi_{nk} \rightarrow \alpha_{nk}}+\upsilon_{\alpha_{nk}\rightarrow \varphi_{nk}}},
\end{equation}
\begin{equation}\label{66}
  \upsilon_{\alpha_{nk}} = \frac{\upsilon_{\varphi_{nk} \rightarrow \alpha_{nk}}\upsilon_{\alpha_{nk}\rightarrow \varphi_{nk}}}{\upsilon_{\varphi_{nk} \rightarrow \alpha_{nk}}+\upsilon_{\alpha_{nk}\rightarrow \varphi_{nk}}}.
\end{equation}
Given posterior belief of $\alpha_{nk}$, the message $I_{\alpha_{nk}\rightarrow f_{tn}}(\alpha_{nk})$ can be updated as,
\begin{align}\label{67}
  \nonumber I_{\alpha_{nk}\rightarrow f_{tn}}(\alpha_{nk}) & = \frac{b(\alpha_{nk})}{I_{f_{tn}\rightarrow \alpha_{nk}}(\alpha_{nk})} \\
  & \propto \mathcal{CN}(\alpha_{nk}; \tau_{\alpha_{nk}\rightarrow f_{tn}}, \upsilon_{\alpha_{nk}\rightarrow f_{tn}}),
\end{align}
where
\begin{equation}\label{68}
  \tau_{\alpha_{nk}\rightarrow f_{tn}} = \upsilon_{\alpha_{nk}\rightarrow f_{tn}}
  \left(\frac{\tau_{\alpha_{nk}}}{\upsilon_{\alpha_{nk}}}-\frac{\tau_{f_{tn}\rightarrow \alpha_{nk}}}{\upsilon_{f_{tn}\rightarrow \alpha_{nk}}}\right),
\end{equation}
\begin{equation}\label{69}
  \upsilon_{\alpha_{nk}\rightarrow f_{tn}} = \left(\frac{1}{\upsilon_{\alpha_{nk}}}-\frac{1}{\upsilon_{f_{tn}\rightarrow \alpha_{nk}}}\right)^{-1}.
\end{equation}

Similar to~\eqref{61}, we can calculate the message passed from function node $\varphi_{nk}$ to variable node $h_{kl}$ as follows:
\begin{equation}\label{70}
  I_{\varphi_{nk}\rightarrow h_{kl}}(h_{kl}) \propto \mathcal{CN}(h_{kl}; \tau_{\varphi_{nk}\rightarrow h_{kl}},\upsilon_{\varphi_{nk}\rightarrow h_{kl}}),
\end{equation}
where
\begin{equation}\label{71}
  \tau_{\varphi_{nk}\rightarrow h_{kl}} = \frac{1}{\mathrm{F}_{nl}}
  (\tau_{\alpha_{nk}\rightarrow \varphi_{nk}}-\sum_{i\neq l}\mathrm{F}_{ni}\tau_{h_{ki}\rightarrow \varphi_{nk}}),
\end{equation}
\begin{equation}\label{72}
  \upsilon_{\varphi_{nk}\rightarrow h_{kl}} = \frac{1}{\mathrm{F}_{nl}^{2}}(\upsilon_{\alpha_{nk}\rightarrow \varphi_{nk}}
  +\sum_{i\neq l}\mathrm{F}_{ni}^{2}\upsilon_{h_{ki}\rightarrow \varphi_{nk}}).
\end{equation}
Further, $I_{h_{kl} \rightarrow q_{kl}}(h_{kl})$ is updated as,
\begin{align}\label{73}
 \nonumber I_{h_{kl} \rightarrow q_{kl}}(h_{kl}) & = \prod_{n} I_{\varphi_{nk} \rightarrow h_{kl}}(h_{kl})\\
   & \propto \mathcal{CN}(h_{kl}; \tau_{h_{kl} \rightarrow q_{kl}},\upsilon_{h_{kl} \rightarrow q_{kl}}),
\end{align}
where the production is through all nonzero subcarriers that user $k$ transmits data and
\begin{equation}\label{74}
  \tau_{h_{kl} \rightarrow q_{kl}} = \upsilon_{h_{kl} \rightarrow q_{kl}}\sum_{n}
  \frac{\tau_{\varphi_{nk} \rightarrow h_{kl}}}{\upsilon_{\varphi_{nk} \rightarrow h_{kl}}},
\end{equation}
\begin{equation}\label{75}
  \upsilon_{h_{kl} \rightarrow q_{kl}} = \left(\sum_{n}\frac{1}{\upsilon_{\varphi_{nk} \rightarrow h_{kl}}}\right)^{-1}.
\end{equation}

For the last part, we discuss the user activity detection loop. Notice that the distribution of active users in a network is sparse and the user activity detection can be treated as a sparse signals learning problem. We extract sparse signals from the received signals based on variational Bayesian inference. To start with, the message $I_{q_{kl}\rightarrow \lambda_{kl}}(\lambda_{kl})$ can be updated as,
\begin{align}\label{76}
 \nonumber I_{q_{kl}\rightarrow \lambda_{kl}}(\lambda_{kl}) & = \exp\{\langle\ln q(h_{kl}|\lambda_{kl})\rangle_{b(h_{kl})}\}\\
    & \propto \mathrm{Gama}(\lambda_{kl};2,|\tau_{h_{kl}}|^{2}+\upsilon_{h_{kl}}),
\end{align}
where $q(h_{kl}|\lambda_{kl})$ is given in~\eqref{10} and the posterior belief $b(h_{kl})$ is given in~\eqref{82}. In a similar approach, the message $I_{q_{kl}\rightarrow h_{kl}}(h_{kl})$ can be calculated as,
\begin{align}\label{77}
 \nonumber I_{q_{kl}\rightarrow h_{kl}}(h_{kl}) & = \exp\{\langle\ln q(h_{kl}|\lambda_{kl})\rangle_{b(\lambda_{kl})}\}\\
    & \propto \mathcal{CN}(h_{kl};0, \tau_{\lambda_{kl}}^{-1}),
\end{align}
where $\tau_{\lambda_{kl}} = \frac{a+1}{b+|\tau_{kl}|^{2}+\upsilon_{kl}}$ and the posterior belief $b(\lambda_{kl})$ is given by,
\begin{align}\label{78}
  \nonumber b(\lambda_{kl}) & = p(\lambda_{kl})I_{q_{kl}\rightarrow \lambda_{kl}}(\lambda_{kl}) \\
  & \propto \mathrm{Gama}(\lambda_{kl}; \hat{a}, \hat{b}),
\end{align}
with $\hat{a} = a+1$ and $\hat{b} = b+|\tau_{h_{kl}}|^{2}+\upsilon_{h_{kl}}$. In~\eqref{77}, $\tau_{\lambda_{kl}}$ serves as the inverse of the channel power, when $\tau_{\lambda_{kl}}^{-1}\rightarrow 0$, the CIR $h_{kl}\rightarrow 0$  as well. Since the inactive users are equivalent to have zero CIRs, an inactive user $k$ is judged when $\sum_{l}{\alpha_{kl}}^{-1} < \delta$ and vice versa, where $\delta$ is a small enough number.

\begin{algorithm}[t]
    \caption{BP-GA-EP message-passing algorithm}
    \label{alg1}
    \begin{algorithmic}[1]
    \REQUIRE $\forall t,n,k,l$,
    $\tau_{h_{kl} \rightarrow \varphi_{nk}}=\hat{h}_{kl}^{ML}, \upsilon_{h_{kl} \rightarrow \varphi_{nk}}=1.0$,
    $\tau_{\alpha_{nk} \rightarrow f_{tn}}=0, \upsilon_{\alpha_{nk} \rightarrow f_{tn}}=1.0$ and
    $\tau_{u_{tnk} \rightarrow f_{tn}}=0, \upsilon_{u_{tnk} \rightarrow f_{tn}}=10^{6}$.
    \ENSURE
        \STATE Update $\tau_{f_{tn} \rightarrow u_{tnk}}, \upsilon_{f_{tn} \rightarrow u_{tnk}}$ via~\eqref{39},~\eqref{40},
               $\forall t,n,k$.
        \STATE Update $I_{f_{tn}\rightarrow \mathbf{x}_{tk}}(x_{tnk})$ and $I_{\mathbf{x}_{tk} \rightarrow f_{tn}}(x_{tnk})$
               via~\eqref{41},~\eqref{42} and~\eqref{43},
               $\forall t,n,k$.
        \STATE Update $\beta(x_{tnk}), \tau_{u_{tnk}}, \upsilon_{u_{tnk}}$ via~\eqref{29}-\eqref{31},
               $\forall t,n,k$.
        \STATE Update $\tau_{u_{tnk} \rightarrow f_{tn}}, \upsilon_{u_{tnk} \rightarrow f_{tn}}$ via~\eqref{36},~\eqref{37},
               $\forall t,n,k$.
        \STATE Update $\beta(x_{tnk}), \tilde{\tau}_{\alpha_{nk}}, \tilde{\upsilon}_{\alpha_{nk}}$ via~\eqref{46}-\eqref{48},
               $\forall t,n,k$.
        \STATE Update $\tau_{f_{tn} \rightarrow \alpha_{nk}}, \upsilon_{f_{tn} \rightarrow \alpha_{nk}}$ via~\eqref{53},~\eqref{54}
               for data symbols and~\eqref{56},~\eqref{57} for pilot signals, $\forall t,n,k$.
        \STATE Update $\tau_{\alpha_{nk} \rightarrow \varphi_{nk}}, \upsilon_{\alpha_{nk} \rightarrow \varphi_{nk}}$ and
               $\tau_{\varphi_{nk} \rightarrow \alpha_{nk}}, \upsilon_{\varphi_{nk} \rightarrow \alpha_{nk}}$ via~\eqref{59},~\eqref{60} and~\eqref{62},~\eqref{63}, respectively, $\forall n,k$.
        \STATE Update $\tau_{\alpha_{nk}}, \upsilon_{\alpha_{nk}}$ via~\eqref{65},~\eqref{66}, $\forall n,k$.
        \STATE Update $\tau_{\alpha_{nk} \rightarrow f_{tn}}, \upsilon_{\alpha_{nk} \rightarrow f_{tn}}$ via~\eqref{68},~\eqref{69}, $\forall t,n,k$.
        \STATE Update $\tau_{\varphi_{nk}\rightarrow h_{kl}}, \upsilon_{\varphi_{nk}\rightarrow h_{kl}}$ via~\eqref{71},~\eqref{72}, $\forall n,k,l$.
        \STATE Update $\tau_{h_{kl}\rightarrow q_{kl}}, \upsilon_{h_{kl}\rightarrow q_{kl}}$ via~\eqref{74},~\eqref{75}, $\forall k,l$.
        \STATE Update $\tau_{h_{kl}}, \upsilon_{h_{kl}}, \tau_{\lambda_{kl}}$ via~\eqref{83},~\eqref{84}, $\forall k,l$.
        \STATE Update $\tau_{h_{kl} \rightarrow \varphi_{nk}}, \upsilon_{h_{kl} \rightarrow \varphi_{nk}}$
               via~\eqref{80},~\eqref{81}, $\forall n,k,l$.
    \end{algorithmic}
\end{algorithm}

In the end, the extrinsic message $I_{h_{kl}\rightarrow \varphi_{nk}}(h_{kl})$ can be updated as,
\begin{align}\label{79}
 \nonumber I_{h_{kl}\rightarrow \varphi_{nk}}(h_{kl}) & = \frac{b(h_{kl})}{I_{\varphi_{nk} \rightarrow h_{kl}}(h_{kl})} \\
   & \propto \mathcal{CN}(h_{kl}; \tau_{h_{kl}\rightarrow \varphi_{nk}}, \upsilon_{h_{kl}\rightarrow \varphi_{nk}}),
\end{align}
where
\begin{equation}\label{80}
  \tau_{h_{kl}\rightarrow \varphi_{nk}} = \upsilon_{h_{kl}\rightarrow \varphi_{nk}}
  \left(\frac{\tau_{h_{kl}}}{\upsilon_{h_{kl}}}-\frac{\tau_{\varphi_{nk} \rightarrow h_{kl}}}{\upsilon_{\varphi_{nk} \rightarrow h_{kl}}}\right),
\end{equation}
\begin{equation}\label{81}
  \upsilon_{h_{kl}\rightarrow \varphi_{nk}} = \left(\frac{1}{\upsilon_{h_{kl}}}-\frac{1}{\upsilon_{\varphi_{nk} \rightarrow h_{kl}}}\right)^{-1}.
\end{equation}
and the posterior belief of $h_{kl}$ is computed with~\eqref{73} and~\eqref{77},
\begin{align}\label{82}
  \nonumber b(h_{kl}) & = I_{q_{kl}\rightarrow h_{kl}}(h_{kl}) I_{h_{kl}\rightarrow q_{kl}}(h_{kl}) \\
   & \propto \mathcal{CN}(h_{kl}; \tau_{h_{kl}}, \upsilon_{h_{kl}}),
\end{align}
where
\begin{equation}\label{83}
  \tau_{h_{kl}} = \frac{\tau_{h_{kl}\rightarrow q_{kl}}}{1+\tau_{\lambda_{kl}}\upsilon_{h_{kl}\rightarrow q_{kl}}},
\end{equation}
\begin{equation}\label{84}
  \upsilon_{h_{kl}} = \frac{\upsilon_{h_{kl}\rightarrow q_{kl}}}{1+\tau_{\lambda_{kl}}\upsilon_{h_{kl}\rightarrow q_{kl}}}.
\end{equation}

The proposed message-passing receiver is referred as BP-GA-EP (belief propagation based Gaussian approximation with expectation propagation) and is summarized in Algorithm.~\ref{alg1}. To deal with the deviated approximation problem in EP, damped update~\cite{31} with scaling factor equals $0.5$ is applied in step $4$. Moreover, for the cases $\upsilon_{u_{tnk} \rightarrow f_{tn}}<0$ and $\upsilon_{f_{tn} \rightarrow \alpha_{nk}} < 0$, a large positive number (e.g., $10^6$) is set to replace the above variances.

\subsection{Complexity Analysis}
The computational analysis of joint detectors is discussed in this subsection. For BP-GA-EP, the computational consumption is dominated by~\eqref{29}-\eqref{34} in data detection loop and~\eqref{46}-\eqref{51} in channel estimation loop. The complexity in calculating $\beta(x_{tnk})$, $\tau_{u_{tnk}}$, and $\upsilon_{u_{tnk}}$ is in the order of $\mathcal{O}(BNMd_{c})$ for each time slot. In parallel with data detection loop, the complexity in calculating $\beta(x_{tnk})$, $\tilde{\tau}_{\alpha_{nk}}$, and $\tilde{\upsilon}_{\alpha_{nk}}$ is also in the order of $\mathcal{O}(BNMd_{c})$. For BP-MF, the interference cancellation structure in~\eqref{23b} dominates the computational consumption in data detection part while in channel estimation part, an updating equation similar to~\eqref{23b} can be obtained. For both parts, the complexity is in the order of $\mathcal{O}(BNMd_{c})$. With the Gaussian approximation, BP-GA has the same complexity with BP-GA-EP for both data detection and channel estimation, i.e., the complexity is in the order of $\mathcal{O}(BNMd_{c})$, see (22) and (23) in~\cite{29}. The joint maximization step in SBL algorithm has a complexity of $\mathcal{O}(M^{K})$, which is prohibitively complexity for massive number of connected users. To reduce the complexity, we resort the EM-decomposition~\cite{34} to simplify the multiuser system into $K$ single user system. Hence, the complexity for data detection in M-step of SBL is given by $\mathcal{O}(KMBd_{v})$. In addition, the complexity for channel estimation involved in E-step is dominated by matrix conversion~\cite{24} which has a cubic complexity $\mathcal{O}(KL^{3})$. Moreover, for BP-MF and SBL, as discussed later in Section IV, an initial estimation of $x_{tnk}$ based MPA is necessary. Thus, there is an additional computational overhead with the order of $\mathcal{O}(BNM^{d_{c}})$ on the initialization step for BP-MF and SBL.

\section{Simulation Results}
\subsection{Experimental Set Up}
\begin{figure*}
  \subfigure[User activity $\lambda=5.0$]{
    \label{Fig.4a} 
    \includegraphics[width=3.3in]{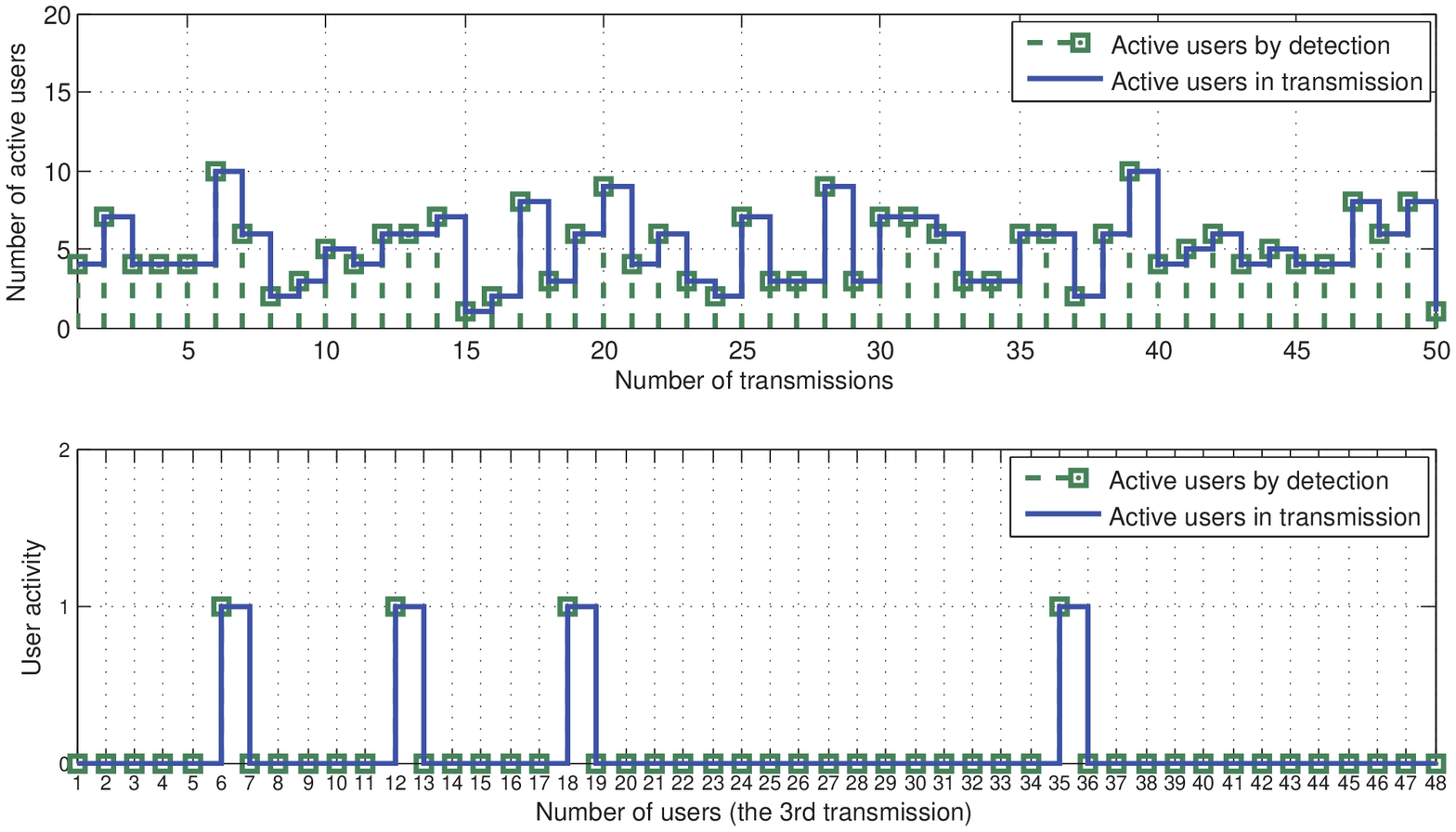}}
  \hspace{0.05in}
  \subfigure[User activity $\lambda=10.0$]{
    \label{Fig.4b} 
    \includegraphics[width=3.3in]{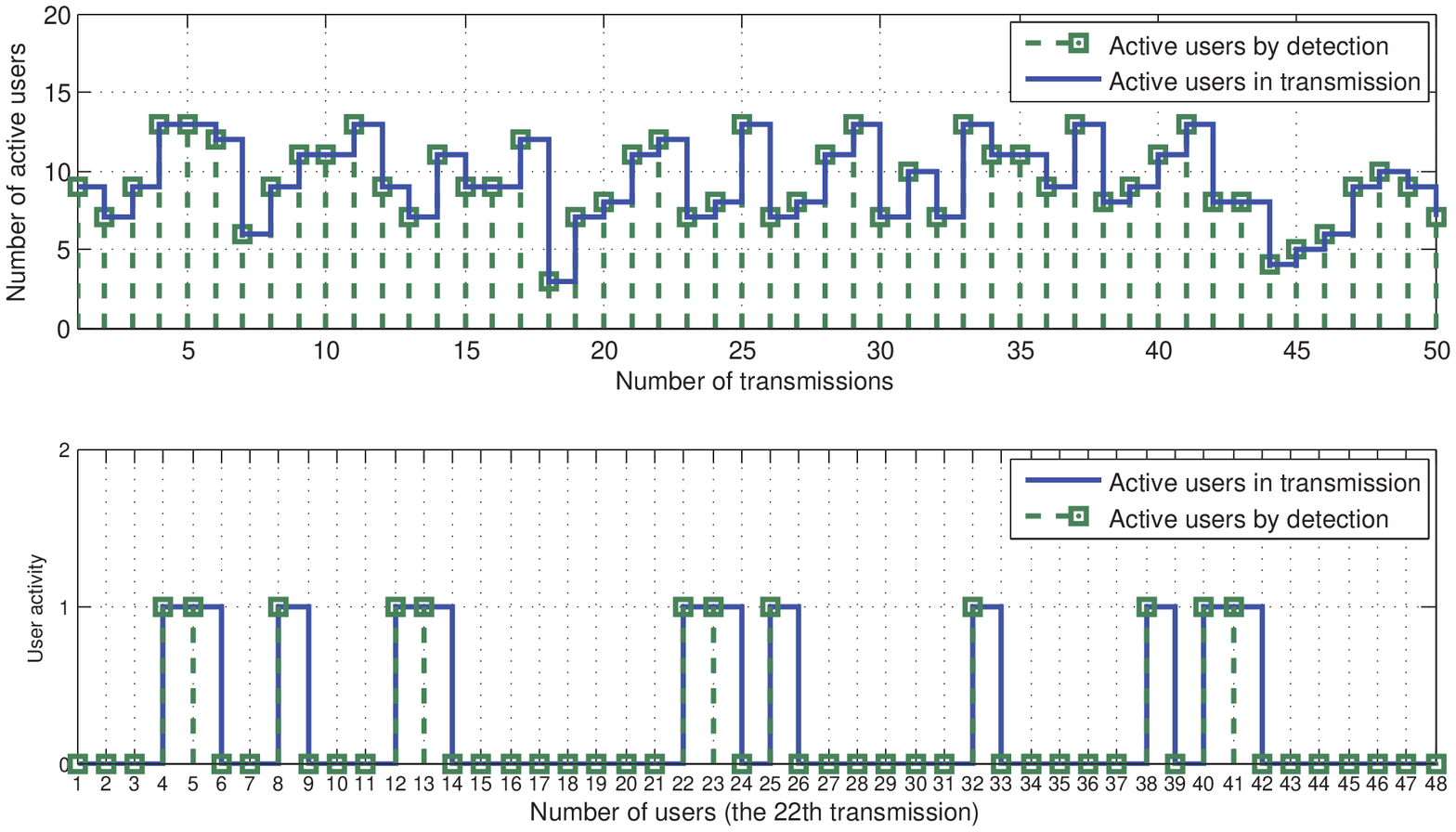}}
  \caption{Active user detection in $E_{b}/N_{0}=3$ dB, $N_{p}=7$, $N_{s}=64$}
  \label{Fig.4} 
\end{figure*}

\begin{figure*}
  \subfigure[User activity $\lambda=5.0$]{
    \label{Fig.5a} 
    \includegraphics[width=3.3in]{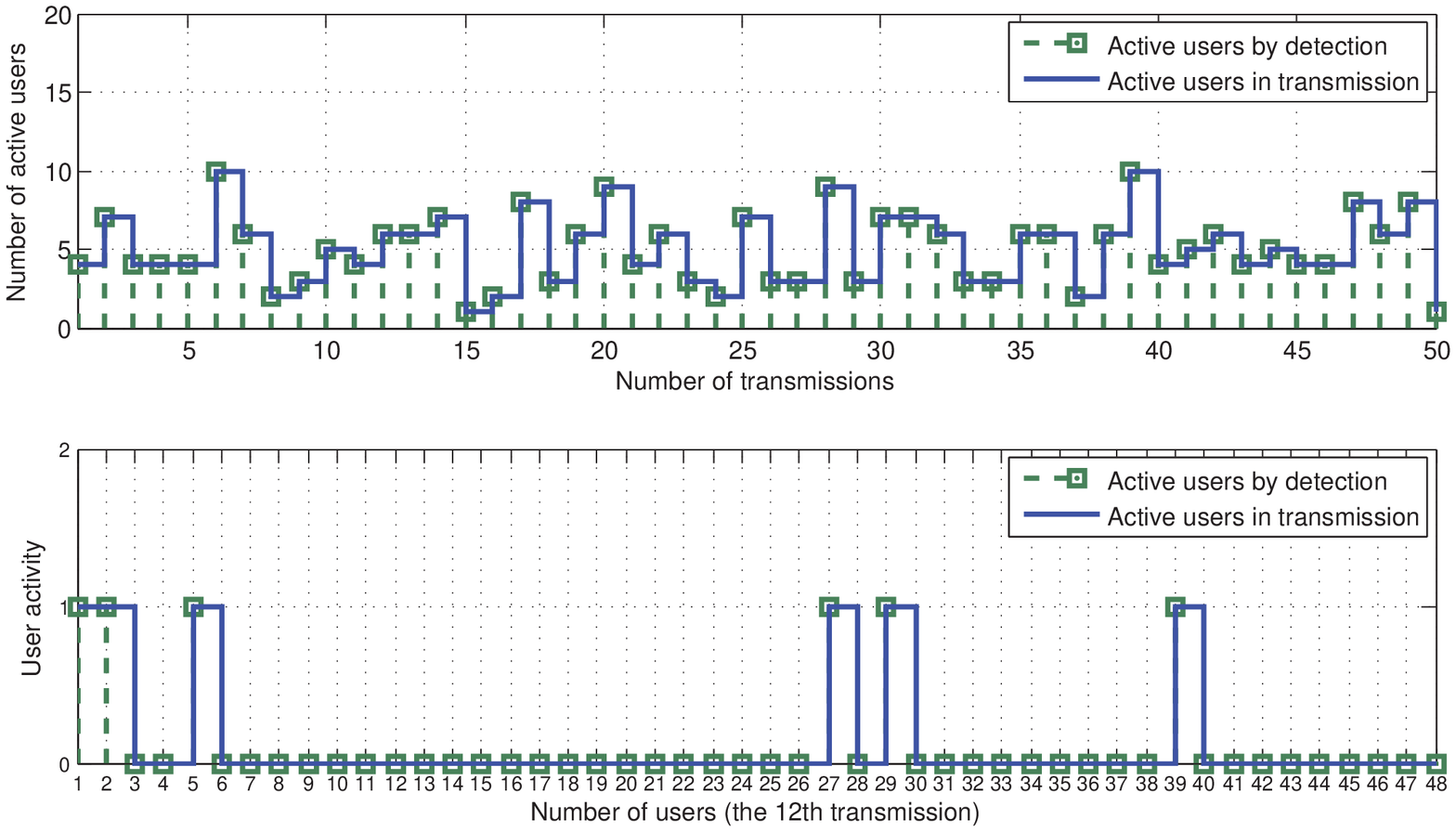}}
  \hspace{0.05in}
  \subfigure[User activity $\lambda=10.0$]{
    \label{Fig.5b} 
    \includegraphics[width=3.3in]{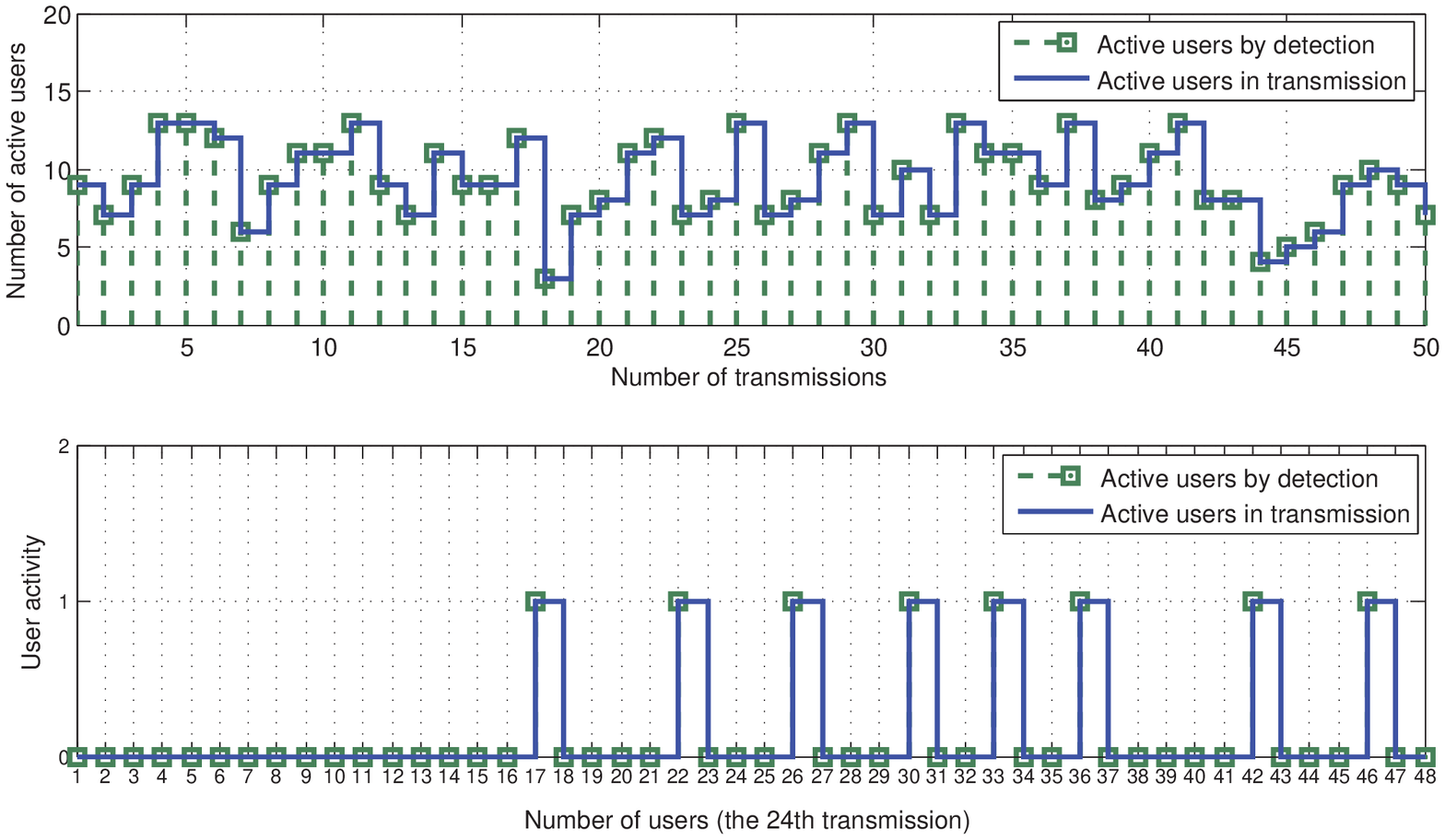}}
  \caption{Active user detection in $E_{b}/N_{0}=5$ dB, $N_{p}=7$, $N_{s}=64$}
  \label{Fig.5} 
\end{figure*}

In this section, the performance of proposed message-passing receiver is evaluated through Monte Carlo simulation. The uplink SCMA system with $B=6, N=24, K=48, M=4, d_{v}=2$, and $d_{c}=4$ is considered in this paper. In each transmission, the active users are generated randomly and we assume the number of simultaneous active users in one transmission follows Poisson distribution parameterized by $\lambda$. The maximum channel order of CIRs for each user is set to $L=6$ with unknown power-delay-profiles follow~\cite{35}. We assume the frequency-selective block-fading channel coefficient $\alpha_{nk}$ remains constant during $N_{s}$ data symbols and $N_{p}$ pilot symbols are multiplexed within one fading block for channel estimation. Zaddoff-Chu sequence designed to be orthogonal in each subcarrier is used as pilot signals. For channel coding, turbo code with rate $R=0.5$ and block size of $9216$ coded bits is generated for each active user.

\subsection{Performance Evaluation}
The accuracy of the proposed grant-free receiver in user activity detection is shown in Figs.~\ref{Fig.4} and~\ref{Fig.5}. We consider different operational SNRs as well as different user activity $\lambda$. In each subfigure, the top one compares the actual number of transmitted users with the number of detected users by receiver in $50$ random transmissions. At the same time, the below one shows the detected active users compared with the actually transmitted users in a random pick up from the above $50$ transmissions. From the figures, we can observe that the proposed receiver has retrieved the sparse distributed active users in the network accurately regardless of the SNRs and user activities. For instance, in Fig.~\ref{Fig.4a}, we evaluate the active user detection in $E_{b}/N_{0}=3$ dB and with the user activity $\lambda = 5$. In the top one, we see the dashed green line that corresponds to the number of detected users is well matched with the solid blue line, which is the number of active transmitted users. In the below one, a random snapshot of the $3_{rd}$ transmission is plotted showing that active users comprising of the $6_{th}$, the $12_{th}$, the $18_{th}$, and the $35_{th}$ user are detected accurately with no leave-out or false-alarm. In deed, the information of user activity is contained in not only pilot signals but also data signals, therefore by exploring both received signals, the active users can be retrieved precisely based on joint detector.

\begin{figure*}[th]
  \subfigure[NMSE comparison]{
    \label{Fig.6a} 
    \includegraphics[width=3.3in]{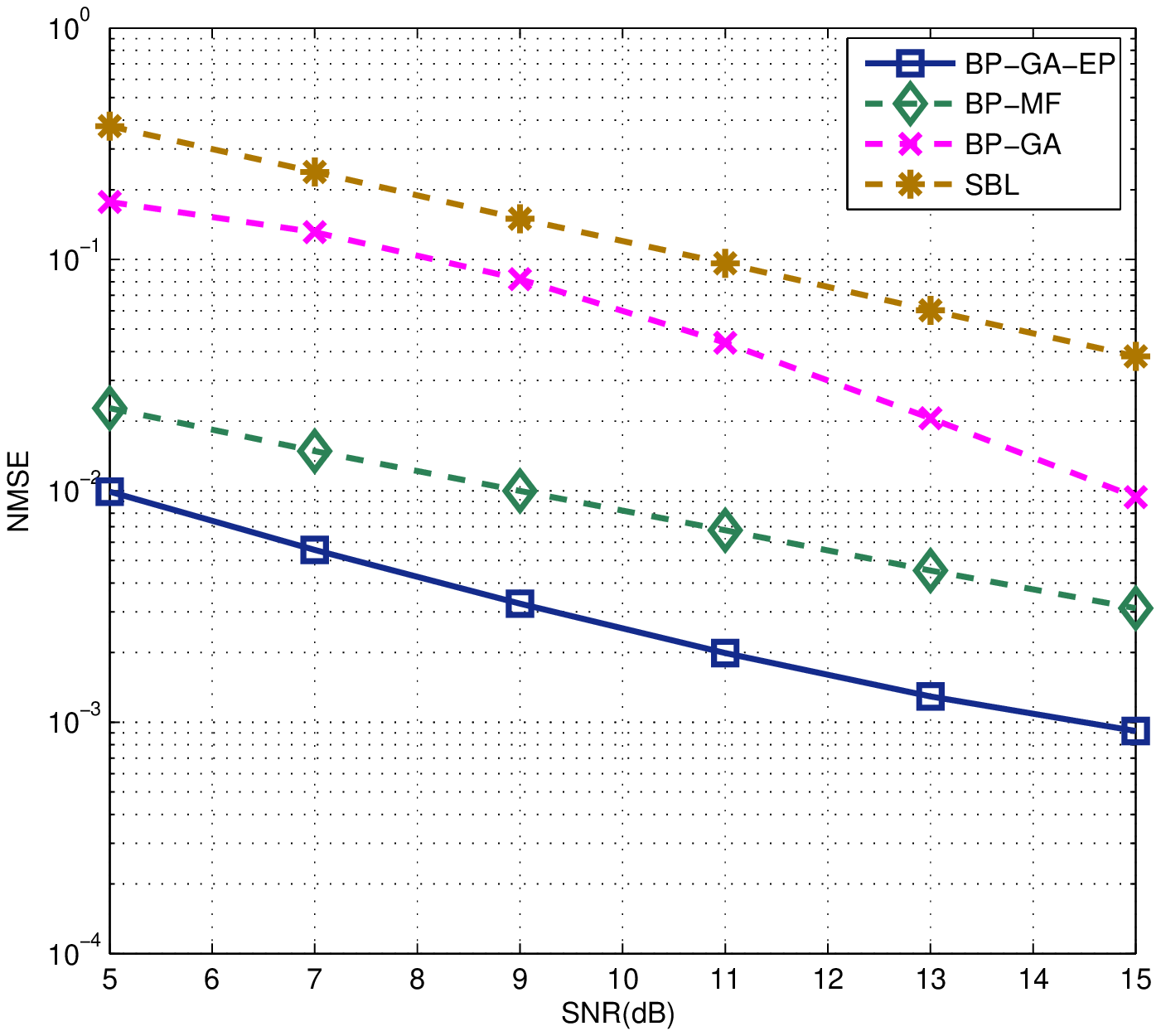}}
  \hspace{0.05in}
  \subfigure[BER comparison]{
    \label{Fig.6b} 
    \includegraphics[width=3.3in]{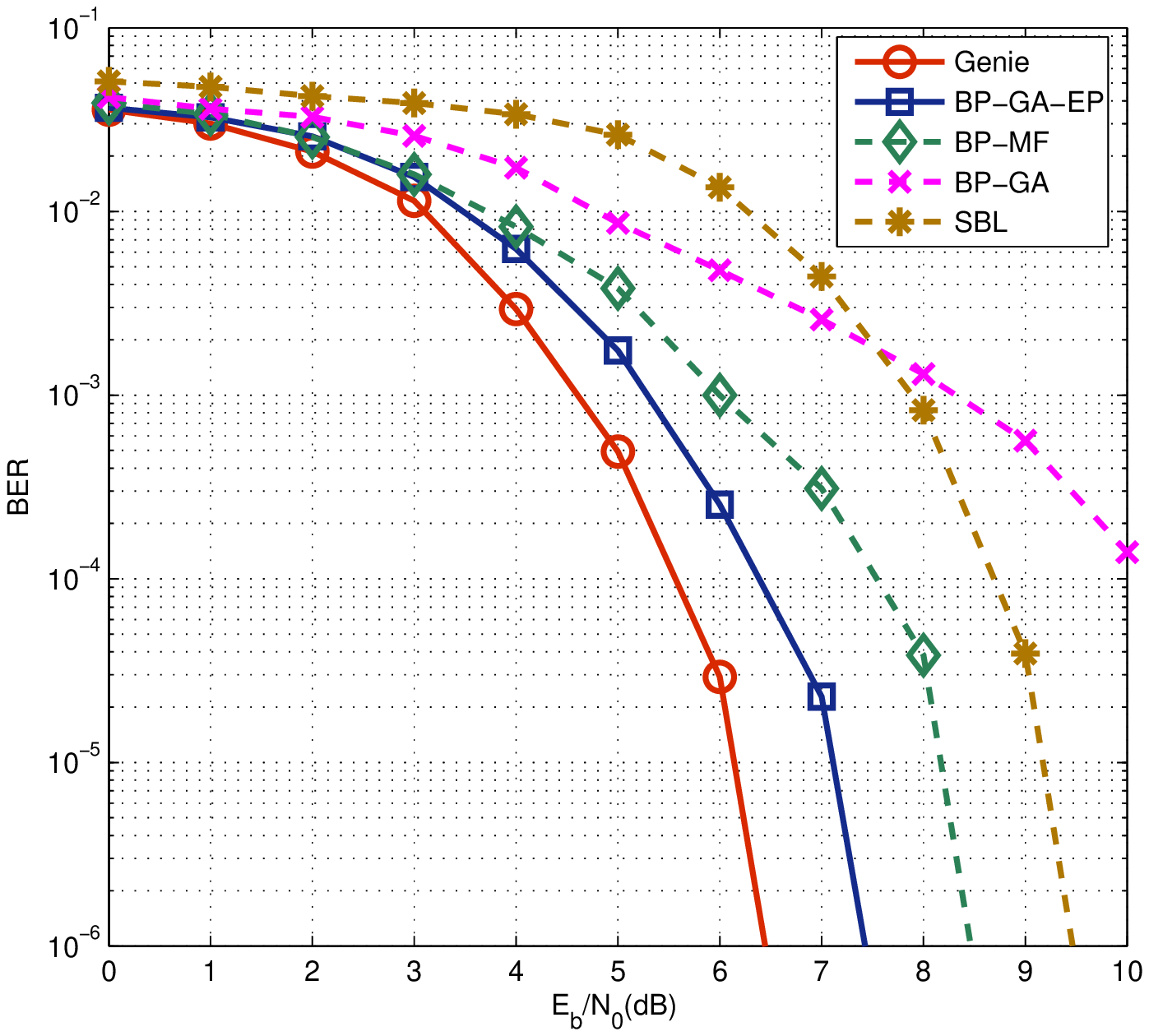}}
  \caption{Performance comparison for iterative joint detectors ($N_{p}=7,N_{s}=64,\lambda=8.0$)}
  \label{Fig.6} 
\end{figure*}

In Fig.~\ref{Fig.6}, the proposed BP-GA-EP algorithm is compared with SBL, BP-MF, and BP-GA in terms of normalized minimum mean square error (NMSE) and bit-error-rate (BER), respectively. As EM-type algorithm and MF can only find a local optimal solution, for SBL and BP-MF, message-passing algorithm is proceeded with a few iterations to get an initial estimate of $x_{tnk}$ through using the ML estimation of CIRs $\mathbf{\hat{h}}_{ML}$ from received pilot signals. For comparison, we also include the genie aided receiver that has a perfect knowledge on both channel state information and user activity in the network.

\begin{figure}
  \centering
  \includegraphics[width=3.5in]{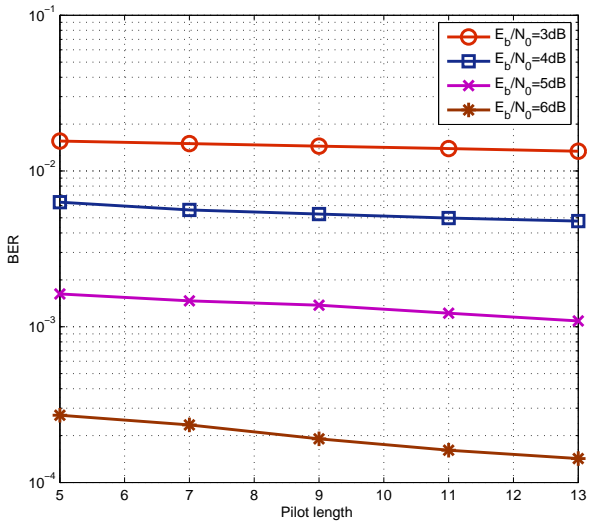}
  \caption{BER performance of BP-GA-EP receiver with different pilot length $N_{p}$, $N_{s}=64$, $\lambda=8.0$}\label{Fig.7}
\end{figure}

\begin{figure}
  \centering
  \includegraphics[width=3.5in]{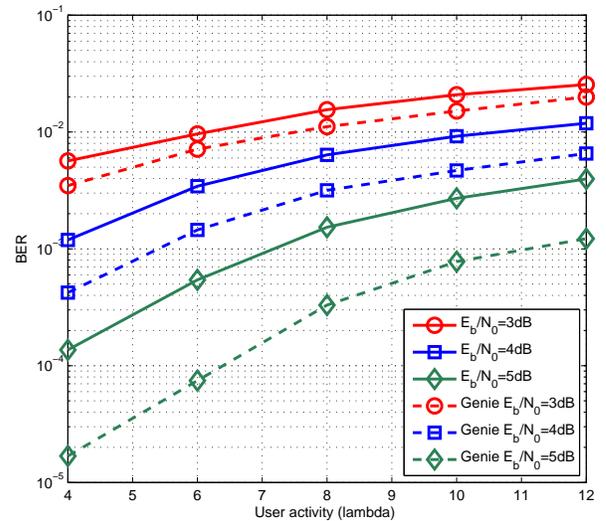}
  \caption{BER performance of BP-GA-EP receiver with different user activity, $N_{p}=7$, $N_{s}=64$}\label{Fig.8}
\end{figure}

From Fig.~\ref{Fig.6}, we observe that compared with genie aided receiver, the current receivers, ranging from SBL, BP-MF to BP-GA all suffer from significant performance loss. For SBL, within each iteration, the EM framework only provides a hard decision on data symbols in M-step (e.g., the maximization (38) in~\cite{24}). The hard decision based on direct maximization is regarded to be inferior to soft decision that offers probability values (e.g, the LLRs) for data symbols. Therefore, SBL has the worst performance on NMSE and BER among the four receivers. For BER, it only performs better than BP-GA in high SNR region. Compared with SBL, the BP-MF receiver can provide soft values for data symbols and thus has an improved performance. However, the estimation is still unconvinced since the LLRs for data symbols only involves the mean values of interferences but omits the variances. Consequently, it operates $2$ dB away from the genie aided receiver in the BER of $10^{-6}$. Based on central-limit theory and moment matching, BP-GA receiver approximates the interferences in each OFDMA subcarrier with Gaussian distribution and can provide the estimations with both mean and covariance. Admittedly, the strategy has gained great success in large scale MIMO-OFDM system that has tens or hundreds interferences. For SCMA, however, the number of collision users in one dimension is limited. As a result, direct Gaussian approximation suffers from a large performance loss. The proposed BP-GA-EP receiver relies on Gaussian approximation as well. Nevertheless, different from BP-GA, the signal for each user is projected into Gaussian distribution individually so that the KL-divergence is minimized. Consequently, the proposed receiver is performance guaranteed. Last but not least, from Fig.~\ref{Fig.6} we observe that the proposed BP-GA-EP algorithm achieves the best performance among four schemes in either NMSE or BER and is only $1$ dB worse compared with the genie aided receiver at BER of $10^{-6}$.

We further evaluate the impact of pilot length on BER performance with proposed receiver at different operational SNRs in Fig.~\ref{Fig.7}. It is shown that the BER is getting slightly improved with the increase of pilot length. In the joint detector, the pilots only provide an ML estimation for CIRs in~\eqref{17}, as a rough initialization for Algorithm~\ref{alg1}. After that, the channel estimation is further refined in joint detection by data symbols. Since the quality of proposed detector depends mainly on the joint estimation step, an increasing length of pilot symbols has minor effects in system performance. In other words, pilot overheads could be reduced by the joint detector.

\begin{figure}
  \centering
  \includegraphics[width=3.5in]{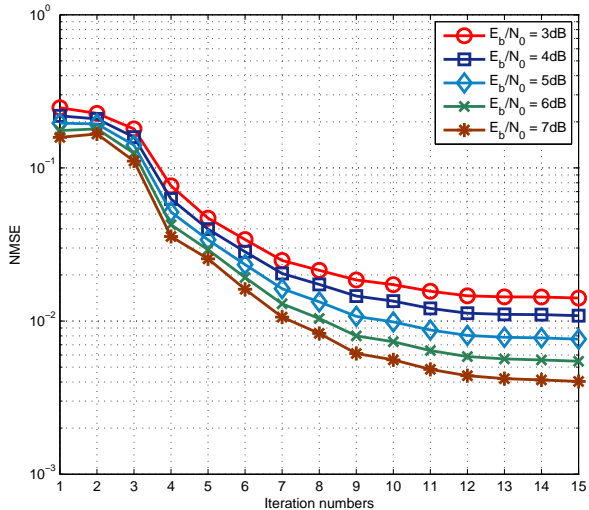}
  \caption{Convergence behaviour of NMSE for BP-GA-EP receiver, $N_{p}=7$, $N_{s}=64$, $\lambda=8.0$}\label{Fig.9}
\end{figure}

\begin{figure}
  \centering
  \includegraphics[width=3.5in]{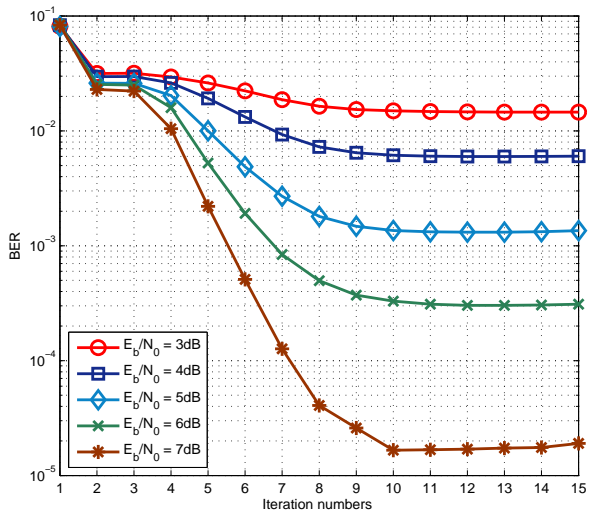}
  \caption{Convergence behaviour of BER for BP-GA-EP receiver, $N_{p}=7$, $N_{s}=64$, $\lambda=8.0$}\label{Fig.10}
\end{figure}

Finally, we demonstrate the impact of user activity on BER performance as well as the convergence behaviour of proposed BP-GA-EP receiver. In Figs.~\ref{Fig.4} and~\ref{Fig.5}, we have shown that the proposed joint receiver can identify active users perfectly in the network regardless of user activity $\lambda$. In Fig.~\ref{Fig.8}, the impact of user activity on BER performance is illustrated. Moreover, the BER of BP-GA-EP increases with the rising of user activity but maintains a constant gap with genie aided curves. It should be noted that with the increase of user activity, the multiuser interference in the system also increases. Therefore, the BER performance of the system deteriorates. In Figs.~\ref{Fig.9} and~\ref{Fig.10}, we demonstrate the convergence behaviour of the proposed BP-GA-EP receiver. Form this figure it can be deduced that the proposed algorithm achieves a stationary point in about $10$ iterations for all considered cases.

\section{Conclusion}
In this paper, we proposed a message-passing receiver for uplink grant-free SCMA. With the lack of information on CIRs and user activity, the proposed receiver performs joint channel estimation, data decoding, and active user detection blindly. By exploring both the received signals from transmitted pilot and data symbols, the joint receiver can identify active users in the network with a high accuracy. For data decoding, the FG of SCMA was formulated and intractable distributions were projected into Gaussian families such that the KL-divergence is minimized. We compared the proposed BP-GA-EP receiver with existing joint detectors along with the genie aided receiver. Through simulations, it was demonstrated that the proposed receiver performs best among current schemes and has about $1$ dB performance gap compared to the genie aided receiver.

\section*{Appendix}
To get the distribution for message $I_{u_{tnk}\rightarrow f_{tn}}(u_{tnk})$ in~\eqref{27}, we first compute the cumulative probability distribution (CDF) which is given by
\begin{align}\label{25}
  \nonumber p(\alpha_{nk}x_{tnk} \leq u_{tnk}) & = \langle p(\alpha_{nk}x_{tnk} \leq u_{tnk}| x_{tnk})\rangle_{I_{\mathbf{x}_{tk}\rightarrow f_{tnk}}(x_{tnk})} \\
  \nonumber & = I_{\mathbf{x}_{tk}\rightarrow f_{tnk}}(x_{tnk}=0)\delta(u_{tnk} \geq 0) \\
            + \sum_{x_{tnk}\neq 0} & I_{\mathbf{x}_{tk}\rightarrow f_{tnk}}(x_{tnk}) p(\alpha_{nk}x_{tnk} \leq u_{tnk} |x_{tnk}),
\end{align}
where $\delta(u_{tnk} \geq 0) = p(u_{tnk} \geq \alpha_{nk}x_{tnk} | x_{tnk} = 0 )$ is an indicator function depending whether $u_{tnk} \geq 0$ or not. Due to this indicator function, the CDF is discontinuous at $u_{tnk} = 0$. Let $F(u_{tnk}) = p(\alpha_{nk}x_{tnk} \leq u_{tnk})$, we have,
\begin{align}\label{26}
  \nonumber p(u_{tnk} = 0) & = \lim_{u_{tnk}\rightarrow 0^{+}}F(u_{tnk}) - \lim_{u_{tnk}\rightarrow 0^{-}}F(u_{tnk})\\
                 & = I_{\mathbf{x}_{tk}\rightarrow f_{tnk}}(x_{tnk}=0).
\end{align}
For $x_{tnk}\neq 0$, the distribution of $u_{tnk}$ can be get in a similar way as in~\cite{33}, Chapter $6$.

\ifCLASSOPTIONcaptionsoff
  \newpage
\fi

\end{document}